\documentclass{emulateapj}
\usepackage{amsmath}
\usepackage{amssymb}
\usepackage{graphicx}

\newcommand{\uvwone}{\textit{uvw1}}
\newcommand{\uvwtwo}{\textit{uvw2}}
\newcommand{\uvmtwo}{\textit{uvm2}}
\newcommand{\Galex}{\textit{GALEX}}
\newcommand{\Swift}{\textit{Swift}}

\defcitealias{menard10}{MSFR10}
\defcitealias{weingartner01}{WD01}
\defcitealias{hk14}{HKB14}
\defcitealias{hoopes5}{H05}
\defcitealias{mathis77}{MRN77}

\begin{document}

\slugcomment{accepted by ApJ}
\shorttitle{UV HALO MORPHOLOGY}
\shortauthors{HODGES-KLUCK, CAFMEYER, \& BREGMAN}

\title{Ultraviolet Halos around Spiral Galaxies. I. Morphology}

\author{Edmund Hodges-Kluck$^{1}$, Julian Cafmeyer$^{1}$ \& Joel N. Bregman$^{1}$}
\altaffiltext{1}{Department of Astronomy, University of Michigan, Ann
  Arbor, MI 48109}
\email{hodgeskl@umich.edu}

\begin{abstract}
We examine ultraviolet halos around a sample of highly inclined
galaxies within 25\,Mpc to measure their morphology and
luminosity. Despite contamination from galactic light scattered into
the wings of the point-spread function, we find that UV halos occur
around each galaxy in our sample. Around most galaxies the halos form
a thick, diffuse disk-like structure, but starburst galaxies with
galactic superwinds have qualitatively different halos that are more
extensive and have filamentary structure. The spatial coincidence of
the UV halos above star-forming regions, the lack of consistent
association with outflows or extraplanar ionized gas, and the strong
correlation between the halo and galaxy UV luminosity suggest that the
UV light is an extragalactic reflection nebula. UV halos may thus
represent $10^6-10^7 M_{\odot}$ of dust within 2--10\,kpc of the disk,
whose properties may change with height in starburst galaxies. 
\end{abstract}


\section{Introduction}
Dust plays an important role in galaxies as a catalyst for instellar
chemistry, a repository of metals, and a heat source for interstellar
gas. The discovery that there is about as much dust outside
of galaxies as within \citep{menard10} suggests that dust also
plays an important role in the disk-halo cycle \citep[however, see][who argue for
a smaller amount of halo dust]{smith16}. Dust is primarily formed in the
atmospheres of evolved stars or supernova remnants, so 
extragalactic dust was likely transported out of the disk by gaseous
flows \citep[e.g.,][]{ferrara91}. Indeed, dusty outflows have been
detected around edge-on galaxies, including filaments seen in
extinction \citep{howk97,howk99} and dust emission at several kpc
above the midplane \citep{mccormick13,melendez15}. Still, it remains
unclear how and when the dust gets into the halo, how long it remains
there, and whether it is altered in the process. These issues
motivate further studies.

Extragalactic dust can be studied through extinction, emission, or
reflection (scattering). In the first case, extraplanar dust lanes can
be seen against the thick disk of stellar light \citep{howk97}, but
above a few kpc from the galaxy midplane the extinction must be measured
toward background continuum sources, such as quasars. Since there are
few quasars behind any one galaxy, \citet{menard10} stacked many
sources to measure the extinction to beyond 1\,Mpc. In contrast,
thermal emission or reflection will only be seen near the galaxy,
since both processes need a nearby light source. 
This makes edge-on systems the ideal laboratories to isolate
extraplanar dust emission and scattered light.

The possibility of extragalactic reflection nebulae (eRN) around
edge-on galaxies was raised
by \citet{ferrara96}, and \citet{hoopes05} reported the discovery of
diffuse ultraviolet (UV) light consistent with an eRN above the disks of
the edge-on starburst galaxies M82 and NGC~253. This conclusion was
reinforced by \citet{hutton14}, and \citet{seon14} detected a UV halo
around NGC~891 that they argue is also an eRN. In \citet[][hereafter,
  HKB14]{hk14} we reported that UV halos are ubiquitous around highly
inclined galaxies, but \citet{sandin15} pointed out that the light
scattered into the wings of the point-spread function (PSF) can
masquerade as a physical halo. \citet{shinn15} modeled several of the
galaxies in \citetalias{hk14} including this effect, and found that
some of the UV halos we had reported are astrophysical while others are
artificial. In principle, eRN are visible at other
wavelengths, and some of the H$\alpha$ attributed to extraplanar
diffuse ionized gas (eDIG) may actually be reflected light
\citep{ferrara96,seon14}. However, UV halos are especially promising
because the scattering cross-section is high in the UV for typical
dust compositions and the sky is relatively dark.

In \citetalias{hk14} we examined galaxies with both \Swift{} UVOT
\citep{roming05} and \Galex{} \citep{martin05} observations,
regardless of physical properties.  In a series of three papers we now
take a more systematic approach to measuring the physical properties
of the UV halos around highly inclined spiral galaxies. In this paper
(Paper~I) we focus on galaxies within about 25\,Mpc to determine the
frequency of physical UV halos, their morphology, and whether they are
eRN. The subsequent papers will present a catalog of total halo fluxes
for a wider sample where morphology cannot necessarily be measured
(Paper~II), and models of the spectral energy distributions (SEDs) for
those galaxies with the highest quality data (Paper~III).

\begin{deluxetable*}{llccccccccccc}
\tablenum{1}
\tabletypesize{\scriptsize}
\tablecaption{Basic Galaxy Parameters}
\tablewidth{0pt}
\tablehead{
\colhead{Name} & \colhead{Type} & \colhead{$T$} & \colhead{$i$} &
\colhead{$M_B$} & \colhead{$d$} & \colhead{$v_{\text{rot}}$} &
\colhead{$E(B-V)$} & \colhead{$B-V$} & \colhead{$m_K$} &
\colhead{$M_*$} & \colhead{$L_{\text{H}\alpha} (10^{40}$} & \colhead{SFR(IR)} \\
               &                &               & \colhead{(deg)} & \colhead{(mag)} & \colhead{(Mpc)} & \colhead{(km s$^{-1}$)} & \colhead{(mag)} & \colhead{(mag)} & \colhead{(mag)} & \colhead{($10^{10} M_{\odot}$)} & \colhead{erg s$^{-1}$)} & \colhead{($M_{\odot}$ yr$^{-1}$)}\\
\colhead{(1)} & \colhead{(2)} & \colhead{(3)} & \colhead{(4)} & \colhead{(5)} & \colhead{(6)} & \colhead{(7)} & \colhead{(8)} & \colhead{(9)} & \colhead{(10)} & \colhead{(11)} & \colhead{(12)} & \colhead{(13)}
} 
\startdata
\cutinhead{Starbursts}					
NGC0253		& SABc	& 5.1	& 90	& -21.23	& 3.25	& 189.8	& 0.019	& 0.69	& 3.772	& 4.36	& 9.59	& 3.97\\
M82			& Scd	& 7.5	& 76.9	& -20.13	& 3.93	& 65.6	& 0.138	& 0.68	& 4.665	& 1.83	& 15.0	& 9.42\\
NGC4631		& SBcd	& 6.6	& 90	& -22.42	& 6.02	& 138.4	& 0.015	& 0.39	& 6.465	& 0.98	& 15.0	& 1.03\\
NGC3628		& Sb	& 3.1	& 79.3	& -21.54	& 10.89	& 215.4	& 0.024	& 0.68	& 6.074	& 4.47	& 4.59	& 3.30\\
NGC4666		& SABc	& 5.1	& 69.6	& -21.10	& 17.28	& 192.9	& 0.022	& 0.64	& 7.055	& 3.10	& 16.8	& 4.83\\
NGC3079		& SBcd	& 6.6	& 90	& -21.56	& 19.28	& 208.4	& 0.01	& 0.53	& 7.262	& 2.91	& 16.9	& 8.17\\
NGC5775		& SBc	& 5.2	& 83.2	& -21.09	& 20.34	& 187.2	& 0.037	& 0.66	& 7.763	& 2.45	& 0.014 & 3.97\\
NGC4388		& Sb	& 2.7	& 90	& -22.13	& 20.5	& 171.2	& 0.029	& 0.57	& 8.004	& 1.55	& 2.35  & 3.16\\
\cutinhead{Normal Spirals}
NGC0055		& SBm	& 8.8	& 90	& -20.09	& 1.94	& 58.7	& 0.012	& 0.33	& 6.249	& 0.09	& 3.42	& 0.06\\
NGC0891		& Sb	& 3.1	& 90	& -20.37	& 9.96	& 212.1	& 0.058	& 0.70	& 5.938	& 3.99	& 5.30	& 2.43\\
NGC2683		& Sb	& 3.0	& 82.8	& -20.42	& 10.08	& 202.6	& 0.029	& 0.75	& 6.328	& 2.98	& 5.56	& 0.41\\
NGC4517		& Sc	& 6.0	& 90	& -21.46	& 10.56	& 139.6	& 0.021	& 0.53	& 7.329	& 0.73	& 		& 0.34\\
NGC4565		& Sb	& 3.3	& 90	& -22.55	& 12.18	& 243.6	& 0.014	& 0.68	& 6.060	& 5.65	& 2.20	& 0.80\\
NGC4096		& SABc	& 5.3	& 80.5	& -20.39	& 12.68	& 144.8	& 0.016	& 0.50	& 7.806	& 0.088	& 5.18	& 0.56\\
NGC4313		& Sab	& 2.1	& 90	& -20.16	& 14.62	& 117.6	& 0.033	& 		& 8.468	& 		& 		& 0.37\\
NGC3623		& Sa	& 1.0	& 90	& -21.02	& 12.77	& 231.2	& 0.022	& 0.78	& 6.066	& 7.17	& 5.76	& 0.38\\
NGC5907		& SABc	& 5.3	& 90	& -21.08	& 16.37	& 226.6	& 0.009	& 0.62	& 6.757	& 5.03	& 14.1	& 2.04\\
NGC4216		& SABb	& 3.0	& 90	& -20.80	& 16.78	& 244	& 0.028	& 0.83	& 6.524	& 6.06	& 		& 0.44\\
NGC4607		& SBbc	& 4.0	& 90	& -20.18	& 17.78	& 98.9	& 0.028	& 0.75	& 9.584	& 0.60	& 1.17	& 0.68\\
NGC4522		& SBc	& 5.9	& 79.2	& -20.91	& 18.29	& 96.4	& 0.018	& 		& 9.8	& 		& 1.67	& 0.40\\
NGC0134		& SABb	& 4.0	& 90	& -21.49	& 18.71	& 220.2	& 0.016	& 0.69	& 6.844	& 5.95	& 		& 4.51\\
NGC4157		& SABb	& 3.3	& 90	& -19.88	& 18.7	& 188.9	& 0.019	& 0.64	& 7.363	& 3.03	& 8.11	& 2.71\\
ESO358-063	& Scd	& 6.9	& 75.6	& -20.34	& 18.98	& 135	& 0.005	& 0.7	& 9.144	& 0.61	& 		& 0.87\\
NGC4217		& Sb	& 3.1	& 81	& -20.08	& 19.37	& 187.6	& 0.015	& 0.75	& 7.582	& 3.67	& 3.08	& 	\\
NGC4330		& Sc	& 6.0	& 78.9	& -20.02	& 19.61	& 115.7	& 0.021	& 		& 9.51	& 		& 		& 0.36\\
NGC3044		& SBc	& 5.5	& 90	& -20.32	& 22.48	& 152.6	& 0.022	& 0.6	& 8.982	& 1.01	& 15.1	& 2.77\\
NGC5170		& Sc	& 4.9	& 90	& -21.13	& 26.8	& 244.7	& 0.07	& 0.7	& 7.628	& 7.46	& 		& 0.79
\enddata
\tablecomments{\label{table.sample} Cols. (1) Name 
(2) Morphological Type (3) Morphological type code (4) Inclination angle (5) $B$ magnitude (6) Distance (7) Circular rotational velocity
(8) Foreground Galactic extinction (9) $B-V$ color (10) $K$-band
  magnitude from the 2MASS Extended Source Catalog \citet{schlafly11}
  (11) Stellar mass using relation from \citet{bell01} 
(12) H$\alpha$ Luminosity (13) Star-formation rate estimated from the \citet{kennicutt98}
relation $\text{SFR} = 4.5\times 10^{-44}
L_{\text{IR}}$\,$M_{\odot}$\,yr$^{-1}$.  $L_{\text{IR}}$ we measure as
defined by \citet{rice88} $L_{\text{IR}} = 5.67\times 10^5 d_{\text{Mpc}}^2
(13.48f_{12} +5.16f_{25}+2.58f_{60}+f_{100})L_{\odot}$, where the
fluxes at 12, 25, 60, and 100\,$\mu$m are in Jy from the IRAS catalog. }
\tablerefs{Values from the NASA/IPAC Extragalactic Database,
  (http://ned.ipac.caltech.edu/), HyperLeda \citet{makarov14} and the
  IRAS catalog (http://irsa.ipac.caltech.edu/Missions/iras.html).}
\end{deluxetable*}

The remainder of this paper is organized as follows: Section~\ref{section.data} 
describes our sample and data sources, and Section~\ref{section.psf} is
focused on correcting for galactic light scattered into the PSF
wings. In Section~\ref{section.morph} we describe the morphology of
the UV halos and compare them to diffuse halos at other wavelengths, and
in Section~\ref{section.fluxes} we use information from the halo
fluxes in each filter to characterize the UV halos. We interpret the
results in Section~\ref{section.discussion} to argue that the UV halos
are likely eRN and connect the UV measurements to extinction and
emission measurements. Section~\ref{section.summary} summarizes the
paper and presents our main conclusions.

\section{Data}
\label{section.data}

\begin{deluxetable*}{lccccccccccccc}
\tablenum{2}
\tabletypesize{\scriptsize}
\tablecaption{Observations}
\tablewidth{0pt}
\tablehead{
\colhead{Name} & \multicolumn{5}{c}{Exposure Time (s)} & \multicolumn{5}{c}{Sensitivity (AB mag)} & \colhead{H$\alpha$} & \colhead{Radio} & \colhead{21cm}\\
\colhead{}     & \colhead{FUV} & \colhead{NUV} & \colhead{\uvwtwo{}} & \colhead{\uvmtwo{}} & \colhead{\uvwone{}} & \colhead{FUV} & \colhead{NUV} & \colhead{\uvwtwo{}} & \colhead{\uvmtwo{}} & \colhead{\uvwone{}} & \colhead{} & \colhead{} & \colhead{} \\
\colhead{(1)} & \colhead{(2)} & \colhead{(3)} & \colhead{(4)} & \colhead{(5)} & \colhead{(6)} & \colhead{(7)} & \colhead{(8)} & \colhead{(9)} & \colhead{(10)} & \colhead{(11)} & \colhead{(12)} & \colhead{(13)} & \colhead{(14)}
} 
\startdata	
\cutinhead{Starbursts}					
NGC0253		& 3281	& 15043	& 		& 		& 		& 25.0	& 26.1	& 		& 		& 		& 	 & 5  & 13 \\				
M82			& 14707	& 29422	& 91603	& 147713& 92788	& 25.9	& 26.4	& 26.8	& 25.7	& 26.5	& 10 & 21 & 24 \\			
NGC4631		& 3147	& 3147	& 1377	& 5474	& 1123	& 24.9	& 25.1	& 24.0	& 24.7	& 23.6	& 10 & 21 & 19 \\			
NGC3628		& 5812	& 17076	& 8095	& 5530	& 3546	& 25.4	& 26.0	& 25.2	& 24.8	& 24.3	& 	 & 21 & 6 \\			
NGC4666		& 5940	& 5940	& 19561	& 19610	& 9858	& 25.4	& 25.3	& 24.7	& 25.7	& 25.1	& 	 & 21 & 18 \\			
NGC3079		& 16108	& 16108	& 8512	& 7252	& 273	& 26.3	& 26.1	& 25.4	& 25.1	& 22.5	& 23 & 21 & 7 \\			
NGC5775		& 2587	& 5770	& 14978	& 23677	& 4180	& 24.6	& 25.4	& 25.6	& 25.8	& 24.5	& 4	 & 21 & 8 \\			
NGC4388		& 2538	& 4993	& 10331	& 7334	& 5331	& 24.6	& 25.3	& 25.4	& 25.0	& 24.7	& 16 & 21 & 3 \\	
\cutinhead{Normal Spirals}
NGC0055		& 29347	& 30691	& 		& 		& 		& 25.5	& 26.5	& 		& 		& 		& 	 & 	  & 20 \\	
NGC0891		& 6047	& 6283	& 15353	& 15200	& 9787	& 25.0	& 25.2	& 25.7	& 25.5	& 25.2	& 16 & 21 & 14 \\				
NGC2683		& 1600	& 1600	& 6407	& 5644	& 4824	& 24.3	& 24.5	& 25.0	& 24.8	& 25.3	& 16 & 21 & 17 \\				
NGC4517		& 1906	& 6403	& 		& 		& 		& 23.8	& 25.5	& 		& 		& 		& 	 & 	  &    \\	
NGC4565		& 12050	& 		& 		& 		& 		& 		& 25.9	& 		& 		& 		& 	 & 21 & 25 \\			
NGC4096		& 1650	& 1650	& 2670	& 3575	& 		& 24.3	& 24.7	& 24.6	& 24.5	& 		& 10 & 	  &    \\				
NGC4313		& 3862	& 3862	& 		& 		& 		& 24.2	& 25.2	& 		& 		& 		& 	 & 	  &    \\	
NGC3623		& 1656	& 1656	& 5300	& 7122	& 3193	& 24.3	& 24.6	& 25.2	& 25.1	& 24.5	& 10 & 5  & 6  \\				
NGC5907		& 1543	& 5423	& 31888	& 45601	& 26482	& 24.2	& 25.5	& 26.3	& 26.3	& 26.0	& 15 & 21 & 1  \\				
NGC4216		& 1672	& 2604	& 		& 		& 		& 24.3	& 24.9	& 		& 		& 		& 16 & 5  & 3  \\				
NGC4607		& 2701	& 6181	& 		& 		& 		& 24.7	& 25.5	& 		& 		& 		& 11 & 	  & 3  \\					
NGC4522		& 2496	& 2496	& 		& 		& 		& 24.7	& 24.9	& 		& 		& 		& 11 & 	  & 9  \\					
NGC0134		& 5998	& 6336	& 4032	& 		& 		& 25.1	& 24.9	& 24.6	& 		& 		& 	 & 	  &    \\	
NGC4157		& 2494	& 1288	& 		& 		& 		& 24.7	& 24.5	& 		& 		& 		& 23 & 21 & 22 \\				
ESO358		& 2456	& 3197	& 		& 		& 		& 24.7	& 25.2	& 		& 		& 		& 	 & 	  &    \\		
NGC4217		& 11774	& 11774	& 		& 		& 		& 25.8	& 25.9	& 		& 		& 		& 15 & 21 & 1  \\				
NGC4330		& 3862	& 3862	& 2429	& 2429	& 2279	& 25.1	& 25.2	& 24.3	& 24.1	& 24.0	& 	 & 	  & 3  \\			
NGC3044		& 1676	& 1676	& 2910	& 2910	& 2585	& 24.3	& 24.6	& 		& 		& 		& 4	 & 21 & 12 \\				
NGC5170		& 1606	& 3307	& 		& 		& 		& 24.0	& 24.9	& 		& 		& 		& 	 & 	  & 2  
\enddata
\tablerefs{\label{table.obs}  H$\alpha$ and radio references: 
(1) \citet{allaert15} (2) \citet{bottema87} (3) \citet{chung09} (4) \citet{collins00} (5) \citet{condon87}
(6) \citet{haynes79} (7) \citet{irwin87} (8) \citet{irwin94} (9) \citet{kenney01} (10) \citet{kennicutt08}
(11) \citet{koopmann01} (12) \citet{lee97} (13) \citet{lucero15} (14) \citet{oosterloo07} (15) \citet{rand96}
(16) \citet{rossa03a} (17) \citet{vollmer16} (18) \citet{walter04} (19) \citet{weliachew78} (20) \citet{westmeier13}
(21) \citet{wiegert15} (22) \citet{yim14} (23) \citet{young96} (24) \citet{yun94} (25) \citet{zschaechner12}
}
\tablecomments{Cols. (2-6) Exposure times (7-11) 3$\sigma$ point-source sensitivity (12-14) References for non-UV data used here.}
\end{deluxetable*}

Our goal is to measure the structure of the diffuse extraplanar UV
light, so we limit the sample to nearby, highly inclined, spiral
galaxies. The initial sample \citep[drawn from the HyperLeda
database\footnote{http://leda.univ-lyon1.fr/};][]{makarov14} includes galaxies within
25\,Mpc with an inclination angle of $i \ge 80^{\circ}$, a Hubble type
of Sa or later, and an absolute $B$-band magnitude of $M_B <
-20$\,mag. The distance limit is based on the need for high $S/N$ in small
regions. We excluded some galaxies for which the inclination angles or types were
obviously incorrect, and for our distance cut
we used the redshift-independent distances from NASA/IPAC
Extragalactic Database\footnote{http://ned.ipac.caltech.edu/} instead
of the HyperLeda values. The initial sample includes 78 galaxies.

Each galaxy must have moderately deep data in at least one
of the \Galex{} or \Swift{} UVOT filters: \Galex{} FUV
($\lambda$1516\AA) and NUV ($\lambda$2267\AA), or the UVOT \uvwone{}
($\lambda$2600\AA), \uvmtwo{} ($\lambda$2246\AA), and \uvwtwo{}
($\lambda$1928\AA) filters. Most of the data come from the \Galex{} or
\Swift{} archives, but we also obtained new UVOT data for several
targets through the \Swift{} program GO~1013198.  Based on our prior
work, we expect a UV halo flux that is about 1\% of the total galaxy flux, so
to accumulate enough $S/N$ we require exposure times
greater than 2-4\,ks in either \Galex{} filter or 5-6\,ks in any UVOT
filter (this excludes the 600\,s exposures from the \Galex{}
all-sky imaging survey). We excluded galaxies that are in regions with strongly variable Galactic cirrus
or are in very crowded fields.  \Swift{} targets must also fit within
the $17\times 17$\,arcmin field of view for accurate background
subtraction. The 26 galaxies that meet this criteria form our working sample,
and the basic properties of these objects are given in
Table~\ref{table.sample}. The UV exposure times and $3\sigma$ point-source
sensitivities are given in Table~\ref{table.obs}. 
The observation IDs used in this paper are given in the Appendix.

Our reliance on archival data means that the sensitivity and filter
coverage vary widely among the galaxies. We also note that a
disproportionate number of galaxies are in the Virgo Cluster, which
has been observed extensively.  Some of the galaxies are included in
published work on UV halos
\citep{hoopes05,hutton14,seon14,hk14,shinn15,baes16}.

We processed the data for each galaxy as described in
\citetalias{hk14}, including cleaning the UVOT and \Galex{} images of
diffuse artifacts in each exposure. We combined the \Swift{} exposures
for each filter into a single, exposure-corrected image. We masked all
the point sources outside the galaxy that were detected at 3$\sigma$
or greater, and we also masked each source detected at 4$\sigma$ or
greater in any one filter in all the remaining filters. This helps to
exclude stars that are close to the galaxy but are not formally
detected at different wavelengths or in shorter observations.

\section{Extended PSF Wings and Halo Light}
\label{section.psf}

\begin{figure*}
\begin{center}
\includegraphics[width=0.95\textwidth]{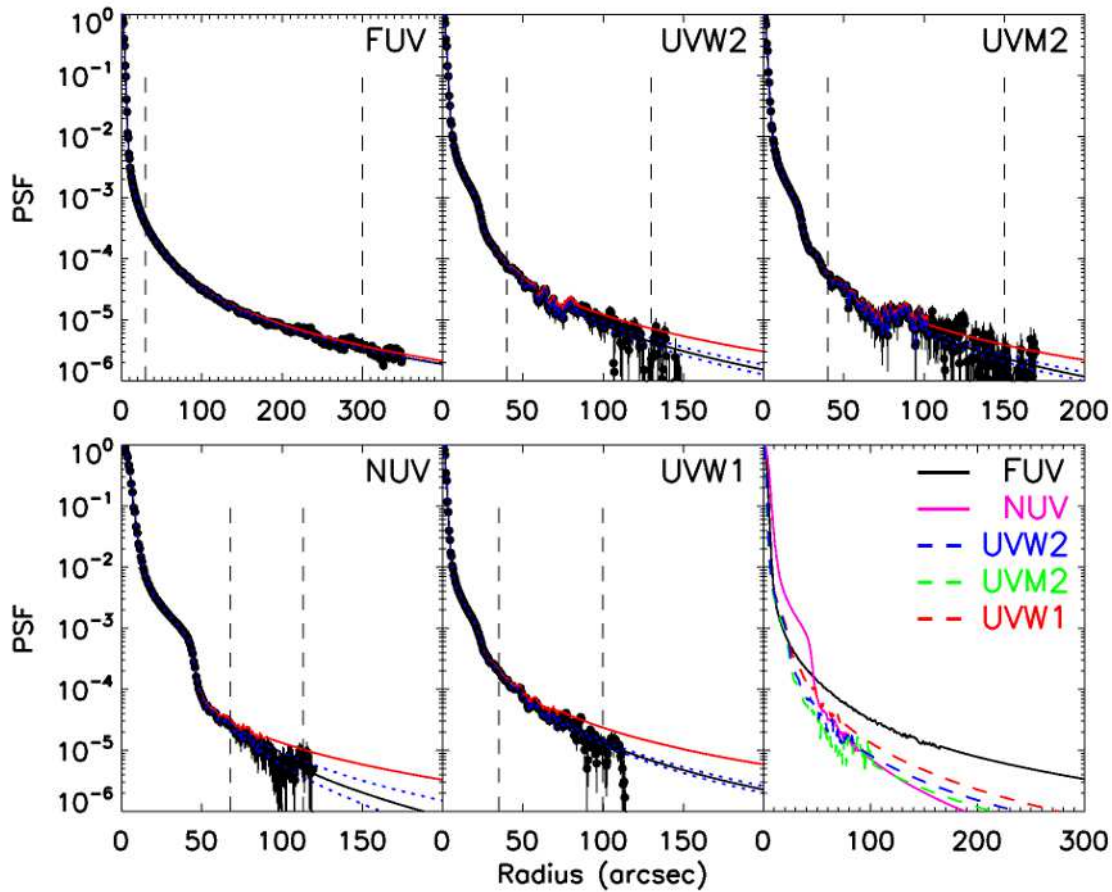}
\caption{\scriptsize Measured, radially-averaged PSF profiles for each
filter (filled circles) and the best-fit profile (black solid
line). The fit is based on points between the vertical dashed lines,
and the profile is extrapolated from the data starting halfway between
these lines. The statistical 1$\sigma$ error bars on the fit are shown as
dashed lines. For comparison, the $r^{-2}$ profile is
shown as a red line. The last panel shows the extrapolated profiles to
a radius of 300\,arcsec for each filter.}
\label{figure.extrapolated_psfs}
\end{center}
\end{figure*}

\citet{sandin15} showed that the diffuse light around edge-on galaxies
consists in part of light from the galaxy image scattered into the
wings of the PSF (the Airy pattern), since the sum of many overlapping
Airy patterns leads to a diffuse glow around the galaxy. To measure and
subtract this component, a PSF must be measured to at least twice the angular
size of the galaxy. Hereafter, we refer to this light as ``PSF-wing 
contamination.'' \citet{shinn15} were the first to address this
issue in the \Galex{} FUV filter, and here we extend their analysis by
measuring and extrapolating the PSFs for each filter in order to
determine the PSF-wing contamination in the measured halo fluxes. 
We want to obtain accurate 2D maps of the halo rather than only a profile 
along the minor axis, so the typical approach of convolving a model with the
PSF is not sufficient.

\subsection{Measured PSFs}

The \Galex{} and \Swift{} calibration teams have provided PSFs
measured to radii of about 60 and 30 arcsec, respectively\footnote{see
  http://www.galex.caltech.edu/researcher/techdoc-ch5.html and
  http://heasarc.gsfc.nasa.gov/docs/heasarc/caldb/swift/docs/uvot/},
but for the galaxies in this work the PSFs need to extend to
200-600\,arcsec.  Extrapolating the wings from
the available PSFs may not be sufficient, so we measured PSFs from
bright point sources to better characterize the wings.

For \Galex{}, we used deep observations of the quasars 3C~273 (30\,ks, tile
GI4\_012003\_3C273) and PKS~$2155-304$ (32\,ks, tile PKS2155m304), which are relatively isolated on the sky
and not bright enough to produce strong ghost images or diffraction
spikes. These quasars are too bright to measure extended PSFs with \Swift{}, 
so we used a combined image of Mrk~501,
which has an effective exposure of 129\,ks in \uvwone{}, 131\,ks in
\uvmtwo{}, and 170\,ks in \uvwtwo{}. The data were processed as
described above and in \citetalias{hk14}, except that we used at least
fifteen sources to compute the astrometric solution for each image and
we rejected images where the source was outside a few arcminutes from
the chip center.
The observation IDs we used are given in the Appendix.

\begin{figure}
\begin{center}
\includegraphics[width=0.5\textwidth]{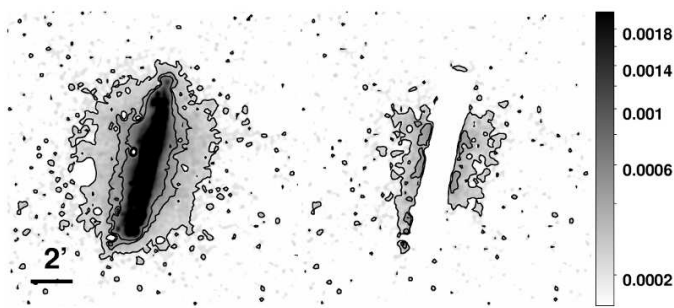}
\caption{\scriptsize \textit{Left}: The FUV image of NGC~3079 before
  correcting for galactic light scattered into the PSF wings. The
  contours are 3, 4, and 5$\sigma$ above background. \textit{Right}:
  The same image after correction for PSF-wing contamination. The
  galaxy image used to compute the correction has been masked.}
\label{figure.before_after}
\end{center}
\end{figure}

We then measured the PSF by constructing an azimuthally symmetric
radial profile to $r>100$\,arcsec. The actual PSF is not azimuthally
symmetric, but simulations show that unless it is very asymmetric (to
an extent that would be noticeable on the image), this approximation
does not noticeably change the measured PSF-wing contamination for any
galaxy with a projected size of about an arcminute or more (even at
smaller sizes, it will not usually make a difference). The background
level is crucial to measuring the small amount of light scattered by
the PSF to large radii, so we masked all point sources detected using
the \Swift{} task {\tt uvotdetect} with a 3$\sigma$ detection
threshold \citep[based on SExtractor;][]{bertin96} other than the
target. Since these sources are \textit{much} fainter than the target,
we assume that masking to the nominal 99\% encircled energy radius is
sufficient.

The measured profiles are shown in
Figure~\ref{figure.extrapolated_psfs}, along with fitted profiles of
the form $A r^{-x} + B$, where $A$, $B$, and $x$ are determined using
least-squares fits to the data between the dashed vertical lines in
Figure~\ref{figure.extrapolated_psfs} (in each plot, $B$ has been
subtracted). For comparison, we also plot a fitted
profile where $x\equiv 2$. The extrapolation based on the best-fit model
begins at the midpoint between the dashed vertical lines, but our
results are insensitive to where we match the extrapolated profile to
the measured one. The FUV is close to $x=2$ at all radii, but the
other filters concentrate light within about 30\,arcsec, with the
subsequent decay being steeper than $x=2$.

The best-fit exponents are given in Table~\ref{table.psf}. The fit
depends strongly on $B$, so we verified that the background level in
the fitting zone was consistent with the background measured outside
the fitting zone.  We also repeated the analysis using a
2$\sigma$ threshold for point source removal and measured $x$ values
that are consistent with the earlier measurements. Finally, we
verified that the PSFs successfully clean the outer regions in the
source images and other images of bright point sources (within
$\sim$10\,arcsec an azimuthally symmetric PSF is a poor match to the data).

There are two caveats to this analysis. First, the PSF becomes
increasingly distorted with increasing distance from the optical axis,
so for images ``far'' from the optical axis the assumption of radial
symmetry is badly wrong (where the critical distance depends on the
instrument).  The PSF is naturally hard to measure to large radii far
from the optical axis, and there are few bright sources that are both
near the chip edge and have very deep \Galex{} or \Swift{} data.
Thus, for galaxies near the chip edge the measured PSFs may not
accurately account for the PSF-wing contamination.

\begin{deluxetable}{lclc}
\tablenum{3}
\tabletypesize{\scriptsize}
\tablecaption{PSF Wing Exponents}
\tablewidth{0pt}
\tablehead{
\colhead{Filter} & \colhead{Exponent} & \colhead{Filter} &
\colhead{Exponent} \\
}
\startdata	
FUV		& $2.05\pm0.01$ & \uvmtwo{} & $2.4\pm0.1$ \\
NUV		& $3.1\pm0.3$   & \uvwone{} & $2.55\pm0.08$ \\
\uvwtwo{}	& $2.4\pm0.1$   & 
\enddata
\tablecomments{\label{table.psf} 
Best-fit exponents ($x$ in $r^{-x}$) and statistical errors from the PSF profile fitting.
}
\end{deluxetable}

Second, the PSF may vary with time \citep[][although the variability
is probably much larger for ground-based instruments]{sandin15}. We
could not make useful tests with \Galex{} data because the bright
sources we used for comparison typically had different PSF cores that
were obviously due to the different positions on the chip. For
\Swift{} we selected all exposures of Mrk~501 in a given filter where
the source was within 1\,arcmin of the chip center, and searched for
variability in the best-fit $x$ between combined images taken in
different months. We found no clear variability, but some low level of
variability beyond 50\,arcsec is possible.
We note that the \Swift{} PSF \textit{core} does vary with time in the
sense that the full-width at half maximum depends on the temperature of
the UVOT focusing rods\footnote{https://heasarc.gsfc.nasa.gov/ftools/caldb/help/uvotapercorr.html}.
However, we did not find any variation in the best-fit exponent for the
PSF wings.

Our results are consistent with the best-fit exponents found in other
sources where the PSF cannot be measured to such large distances, so we 
believe that the fits are reasonably accurate. 

\subsection{Estimating the PSF-wing Contamination}

We estimated the PSF-wing contamination for each image by creating a
model using the input image and the extrapolated PSF models, using the
best-fit exponent for each filter (Table~\ref{table.psf}). We first
clipped the input image to the region that just contains the galaxy.
We did not use an objectively defined region because of the varied
disk morphologies and inclinations in our sample, but the galaxy
regions are conservative in that they extend to at least 2\,kpc above
the midplane. We then convolved the clipped image with a PSF model
that was extrapolated to at least twice the angular diameter of the
galaxy.

The resulting image shows the amplitude of the PSF-wing light at each
position around the galaxy, but the total flux in the model image is
too low and the light is too smeared out because we do not deconvolve
the galaxy image before convolving it with the extrapolated PSF (the
galaxy is too complex for accurate deconvolution). The former effect
is corrected by multiplying the fake image by the ratio of the flux in
the galaxy in the original and fake images (this is not an exact
correction, but simulations of exponential disk models show that it
underestimates the PSF-wing flux by much less than the statistical
uncertainties for our sample). The latter effect cannot be easily
corrected, but we find (again using simulations of exponential disk
models) that the most severe impact is within about 10 pixels of
bright regions in each galaxy, and in practice the galaxy regions we
define extend at least this far from bright clumps. The combination of
these issues leads to an overestimate of the PSF-wing contamination by
about 10\% immediately adjacent to the galaxy, and which
quickly declines with height. We also investigated the error due to
using a symmetric PSF in the FUV, where the PSF is measured to over
100\,arcsec.  We convolved model images with angular sizes smaller
than this with our model PSFs and the measured PSF. This leads to
differences of 0-3\% in the PSF-wing light measured around the
galaxy. On the scale of the whole halo, these effects are very modest.

The model image can then be used in conjunction with the original
image to measure the PSF-wing contamination at any point in the halo,
and we also subtract the model from the original to create corrected
halo images. The PSF-wing contamination differs for each galaxy and
filter, but it ranges from 1-90\% of the total extraplanar flux in our
sample. The contamination is generally larger in the FUV than the
other filters. The importance of the correction is illustrated in
Figure~\ref{figure.before_after} for NGC~3079, where we show FUV
images before and after subtracting the model image. We verified the
\citet{sandin15} proposition that convolving the galaxy image with a
PSF extrapolated to more than twice the angular size of the galaxy
does not significantly alter the PSF-wing contamination.  In
Section~\ref{section.fluxes} we assess the reliability of our PSF-wing
subtraction.

\begin{figure*}
\begin{center}
\includegraphics[width=0.95\textwidth]{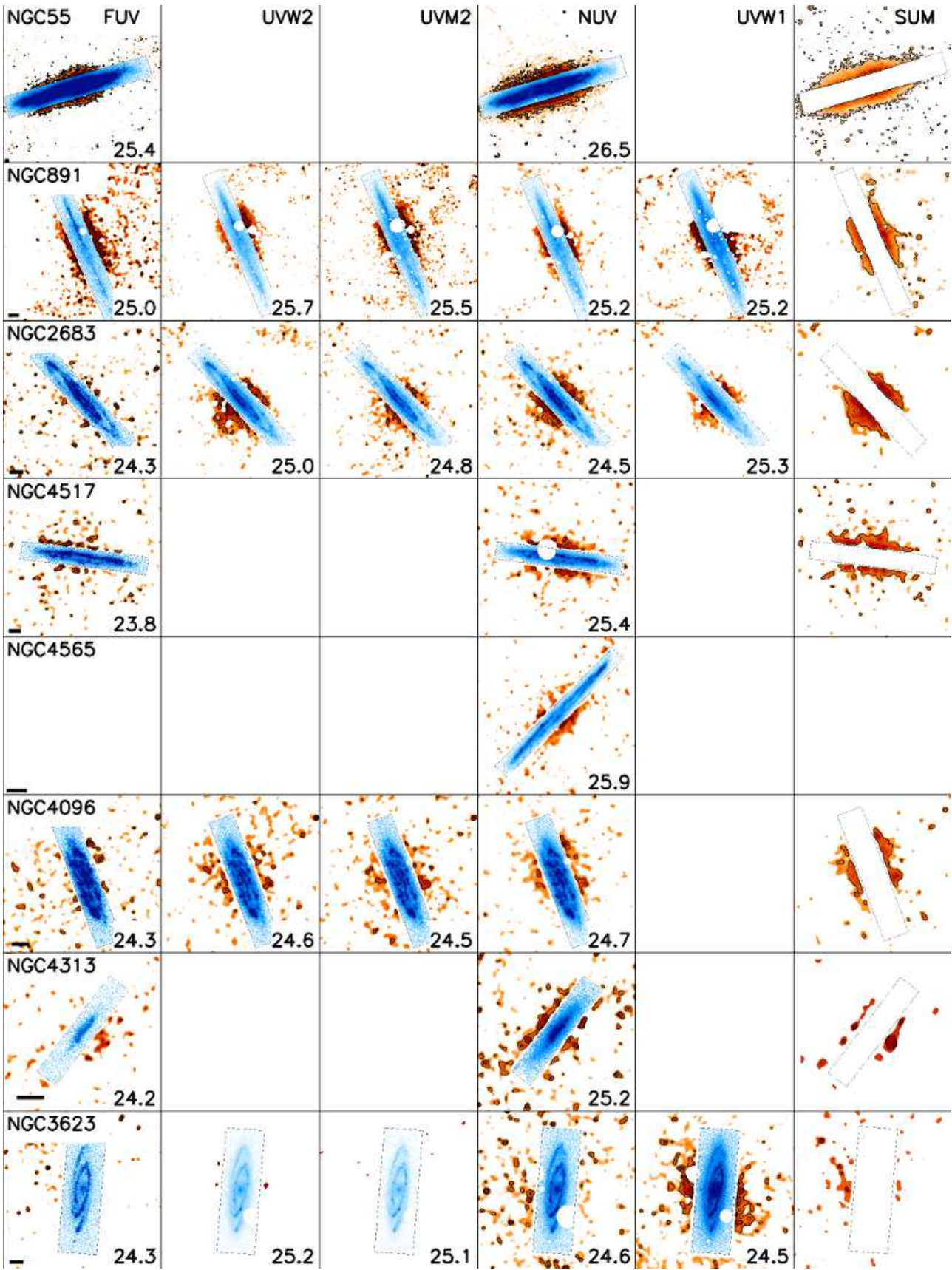}
\caption{\scriptsize UV halos (orange) around normal galaxies in the
  \Galex{} and UVOT filters, shown with increasing wavelength from
  left to right. The final column shows the combined data, referenced to
  the \Galex{} resolution. 
  The orange maps have been corrected for the
  PSF-wing contamination from the galaxy (image shown in blue) and
  point sources have been masked. For each filter, the
  contours are the 2, 4, 8, and 16$\sigma$ contours above the
  background. A 2\,arcmin line is shown at the bottom left of each
  row, and the 3$\sigma$ point-source sensitivity for the map is shown
  at the bottom right of each panel. The composite image only shows the 3$\sigma$
  contour.
  \label{figure.gallery_normal}
}
\end{center}
\end{figure*}
\setcounter{figure}{2}

\begin{figure*}
\begin{center}
\includegraphics[width=0.95\textwidth]{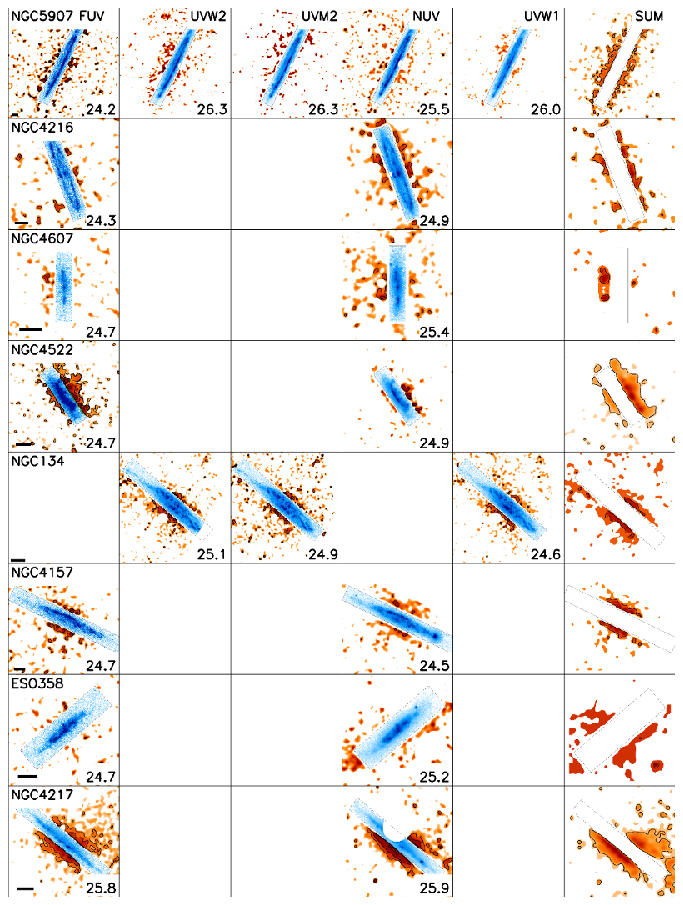}
\caption{\scriptsize continued}
\end{center}
\end{figure*}
\setcounter{figure}{2}

\begin{figure*}
\begin{center}
\includegraphics[width=0.95\textwidth]{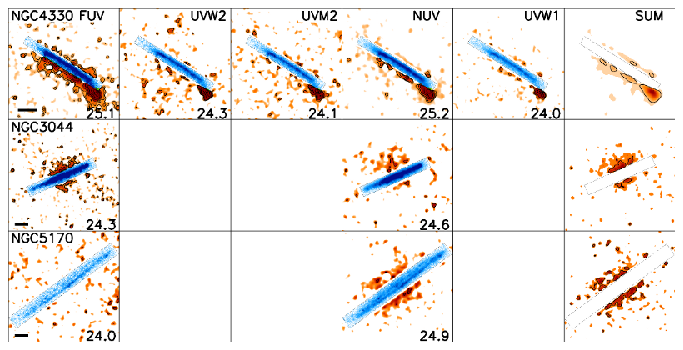}
\caption{\scriptsize continued}
\end{center}
\end{figure*}

\section{Extraplanar Ultraviolet Morphology}
\label{section.morph}

In this section we describe the qualitative structure of the diffuse
extraplanar UV light in our sample and compare it to the diffuse
structure seen at other wavelengths where the source of the light is
basically understood. The initial work on UV halos \citep{hoopes05}
focused on galaxies with galactic superwinds, and in this section we
divide our sample into galaxies with known winds (starburst galaxies)
and those without (normal galaxies).

\subsection{Existence of UV Halos}

\begin{figure*}
\begin{center}
\includegraphics[width=0.95\textwidth]{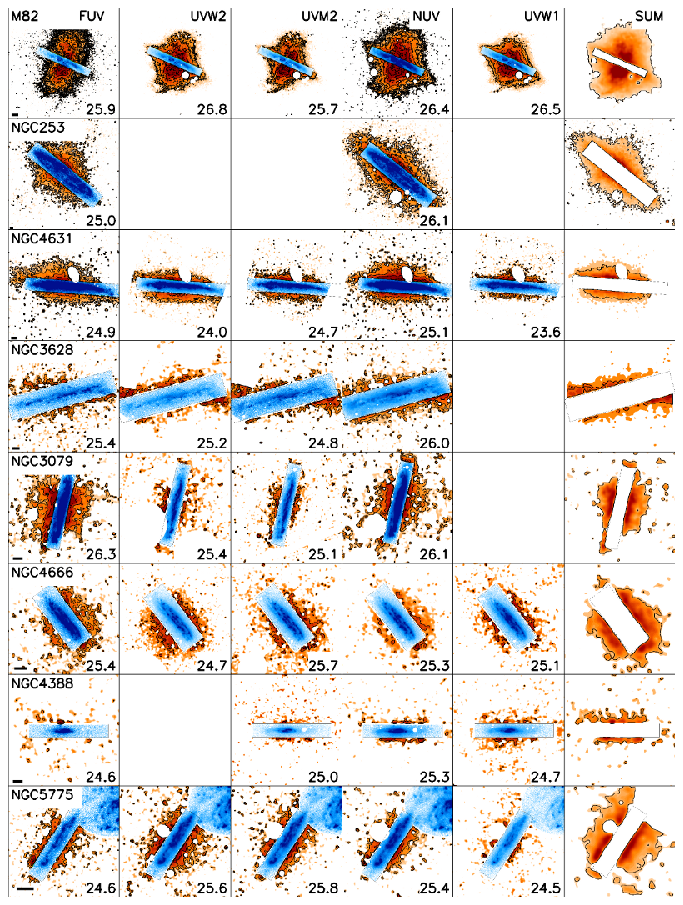}
\caption{\scriptsize UV halos (orange) around starburst galaxies. The
  format is the same as in Figure~\ref{figure.gallery_normal}.
\label{figure.gallery_starburst}
}
\end{center}
\end{figure*}

We use UV images where the PSF-wing contamination has been subtracted.
Figures~\ref{figure.gallery_normal} and
\ref{figure.gallery_starburst} show the images in each filter 
for the normal and starburst galaxies,
respectively. In these images, the galaxy region used to estimate the
PSF-wing contamination is masked, so we superimpose the unsmoothed
image of the galaxy for that filter in this region (in blue). Outside
of this region, point sources have been masked, the image has been
smoothed with a Gaussian kernel (with $\sigma=3-8$\,pixels, depending
on the angular size of the galaxy), and the image has been clipped to
1$\sigma$ above the background. The contours show the 2, 4, 8, and
16$\sigma$ level above the background. For each image we also list the
3$\sigma$ point-source sensitivity in AB magnitudes \citep{oke83},
which shows the differences in depth between filters.

Although the UVOT and \Galex{} images have different PSFs, artifacts,
and systematic uncertainties, the morphology of the residual
extraplanar UV light is similar between filters for most galaxies, so
UV halos appear to be physical phenomena. Thus, in the final columns
of Figures~\ref{figure.gallery_normal} and
\ref{figure.gallery_starburst}
we show a composite image for each galaxy made by summing each filter
(referenced to the effective area of the \uvmtwo{} filter). These
images have roughly twice the $S/N$ of the images in each filter, so
they highlight the extent of the halo. Remarkably, every galaxy in the
sample has a UV halo extending at least to 6\,kpc above the
midplane. Considering that our selection criteria were distance,
luminosity, and inclination, this suggests that UV halos are a common
feature of luminous spiral galaxies.

There are some differences between filters for several galaxies that
may be artificial. For example, in
the \uvwtwo{} images of NGC~4666 and NGC~5775
(Figure~\ref{figure.gallery_starburst}), the halo appears to be more
prominent than in the \uvmtwo{} or FUV images, even though the
\uvmtwo{} has a higher sensitivity. This may indicate residual light
in the central 2\,arcmin halo ring in the large-scale \uvwtwo{}
scattered-light artifact \citepalias{hk14}, which could mimic a halo in an image
centered on the galaxy. The smaller field of view of the UVOT can also
lead to differences: in the relatively shallow UVOT images of NGC~4631
(Figure~\ref{figure.gallery_starburst}), the halo is both less
prominent and more symmetric than in the \Galex{} images. This
probably results from the ``background'' regions in the UVOT being
inside the UV halo itself, a scenario that is supported by the \Galex{} images. 
 In other cases, differences
between the morphology in each filter likely indicate physical
differences. For example, in M82 and NGC~253
(Figure~\ref{figure.gallery_starburst}), the NUV halo appears more
like a thick disk than the FUV halo, which is more filamentary in
nature. This morphology cannot be explained by PSF-wing contamination
or filter artifacts.

We investigated the contribution of stellar clusters or background
sources to the diffuse light in those galaxies with \textit{HST}
observations in the $U$ band or bluer. For these objects, we
identified point-like sources in the \textit{HST} maps that were not already
identified and removed in our maps and predicted a \Swift{} or
\Galex{} flux for each, assuming that each source has the spectrum of
a B star. We then compared the flux to the total diffuse UV flux, and
found that it can explain at most a few percent of the light. The
exception that proves the rule is NGC~4522, where the Virgo cluster is
stripping the ISM \citep{kenney99} and there are known stellar clusters in
the halo. In this case, most of the flux in the FUV comes from these
clusters. Even NGC~4522 has a diffuse component, however, so we
conclude that all of these galaxies have true diffuse UV halos.

\subsection{Ultraviolet Morphology}

There are three qualitative morphological metrics we use to compare
the UV halos in our sample: filamentary structure, concentration of
the UV light (both in the vertical direction and along the disk), and
symmetry across the midplane. In general, these are consistent among
filters for a given galaxy.

\paragraph{Filaments}
Filaments are most easily visible in the FUV, and they are 
seen in most starburst galaxies (M82, NGC~253, NGC~3079,
NGC~3628, NGC~4666, and NGC~5775), as well as two normal galaxies
(NGC~4522, NGC~4330). The filaments in the latter two are associated
with gas stripped by ram pressure in the Virgo cluster. We cannot rule
out the presence of filaments in non-stripped normal galaxies, but
those galaxies with high $S/N$ images do not have filaments. 

\begin{figure*}
\begin{center}
\includegraphics[width=0.95\textwidth]{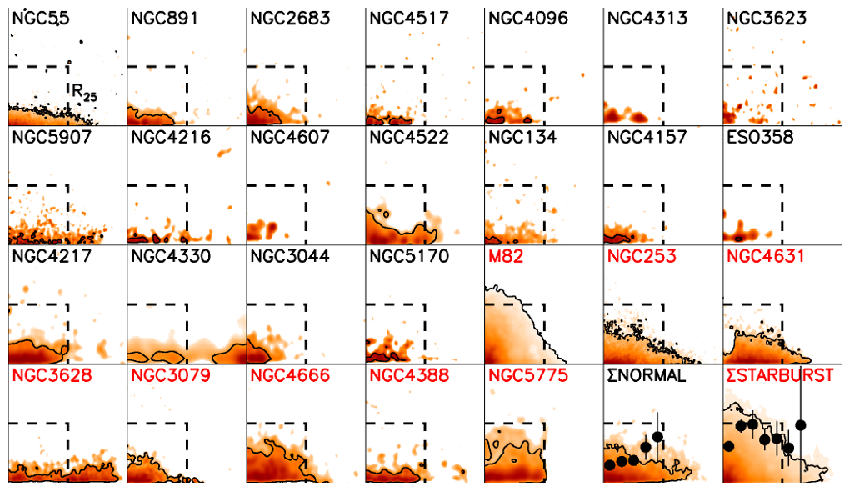}
\caption{\scriptsize Combined images from each filter stacked in the
  first quadrant and shown relative to the optical radius, $R_{25}$. 
  The contour represents the combined 3$\sigma$
  threshold above background, and the dimensions of each image are $2
  R_{25} \times 2 R_{25}$. The dashed lines show $R_{25}$. Starburst
  galaxies are labeled with red names. The final two panels show the
  combined normal and starburst galaxies, respectively, with scale heights
  measured as a function of galactocentric radius shown as points. The
  ``normal'' stack excludes NGC~4330 and NGC~4522, which are being stripped by
  ram pressure.
  The dip at the center of NGC~4631 is due to masking a satellite galaxy, but
  it is accounted for in the starburst stack and does not cause the central 
  dip in the scale height there.
}
\label{figure.quadrants}
\end{center}
\end{figure*}

\paragraph{Concentration}
The UV halos differ in extent, both in galactocentric radius and
projected height. For example, in NGC~891, NGC~2683, and NGC~3044 most
of the light is concentrated near the center of the galaxy
(Figure~\ref{figure.gallery_normal}), whereas in NGC~4216 or NGC~5907
the surface brightness is comparable across most of the disk. We
examined the degree of concentration by dividing the UV halo at the
midplane and projected galactic center into four quadrants, which were
summed. Figure~\ref{figure.quadrants} shows the resulting images.
Each panel has dimensions of $2 R_{25} \times 2 R_{25}$, where
we use the HyperLeda values for $R_{25}$. 
The final two panels show the
sum of all of the normal and starburst galaxies, respectively, where the
images were reprojected and stacked on a common axis relative to $R_{25}$.
We excluded the stripped galaxies NGC~4330 and NGC~4522 from this stack.

There are two aspects to the concentration: the extent of the light
(as measured by the 3$\sigma$ contour in
Figure~\ref{figure.quadrants}) and its distribution. Among the UV
halos of normal
galaxies, 6/18 extend in the radial direction to $R_{25}$ and only one
(NGC~4522) extends beyond $0.5 R_{25}$ in the vertical direction. In
contrast, all of the starburst halos (8/8) extend to $R_{25}$ in the
radial direction, and 6/8 extend to $0.5 R_{25}$ in the vertical
direction. In terms of distribution, for most normal galaxies the UV
light is brightest and most vertically extended above the galactic
center, with a modest decline with increasing galactocentric
radius. This is consistent with a disk-like structure. The starburst
galaxies other than NGC~4631 are also 
less concentrated and filamentary structure at larger heights is
superimposed. The similarity in sensitivity and distance between several of
the starburst and normal galaxies (e.g., NGC~3079 and NGC~5907)
indicates that UV halos are intrinsically brighter and more
morphologically complex around starburst galaxies. 

When compared to near-IR images, the radial extent of the UV halo
along the major axis (relative to $R_{25}$) is unrelated to the
prominence of the bulge, so it is unlikely that the radially
concentrated UV halos are merely light from the bulge outskirts. The
radial or vertical concentration is also unrelated (within each
subsample) to the total halo luminosity. The radial concentration also
appears to be unrelated to the vertical concentration, and the normal
galaxies appear to span a narrow range in vertical extent below $R_{25}$.

\paragraph{Bilateral Symmetry}
We determined the degree of bilateral symmetry by folding images
across the midplane. Most galaxies 
show some asymmetry, although in most cases the visible
structure is symmetric while the flux is not, which could be a
function of the inclination angle. Among the normal galaxies the
bilateral asymmetry is only pronounced for NGC~134, NGC~4330, and
NGC~4522, which are experiencing ram-pressure stripping
\citep{kenney99,abramson11}. Among the starburst galaxies there is
asymmetry in the filamentary structures, but overall the halos are 
nearly symmetric. An exception is NGC~4631, in which the northern part of
the halo is somewhat more extended. NGC~3628 is not bilaterally
symmetric, but its halo appears to be warped; if we invert the image
along the axis of the midplane, it becomes nearly symmetric. A hint of
a similar warp can be seen in the combined image of NGC~4388.

\subsection{FUV$-$NUV Color}

\begin{figure}
\begin{center}
\includegraphics[width=0.5\textwidth]{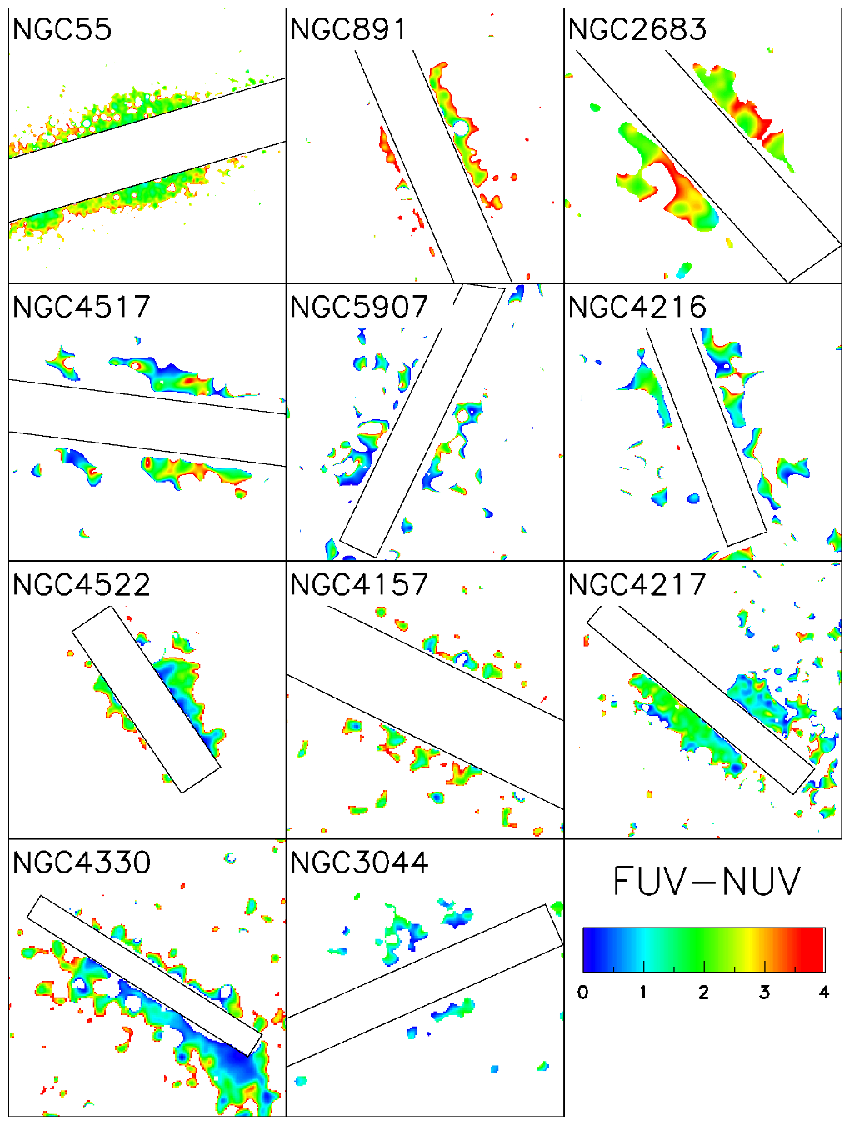}
\caption{\scriptsize The FUV$-$NUV color (per square arcsec) for normal
  galaxies. The FUV and NUV images used to create these maps were
  smoothed and clipped at the mutual 2$\sigma$ level (i.e., where
  emission is detected at 2$\sigma$ above background in each
  image). The \Galex{} pixel size is 1.5\,arcsec.}
\label{figure.fuvnuv_normal}
\end{center}
\end{figure}

\begin{figure}
\begin{center}
\includegraphics[width=0.5\textwidth]{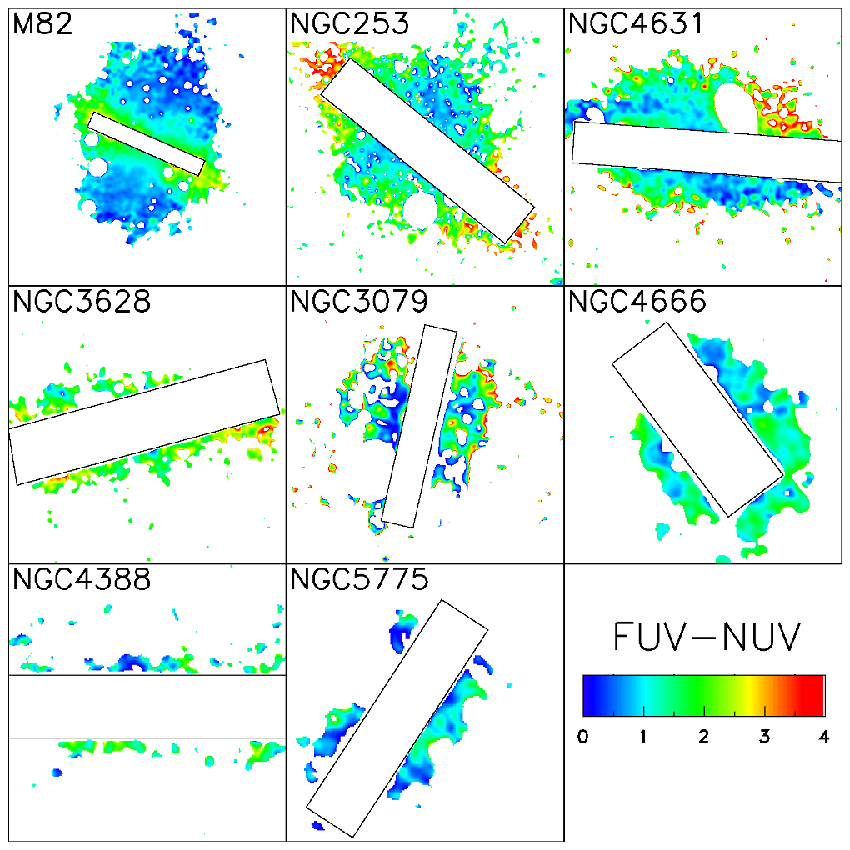}
\caption{\scriptsize The FUV$-$NUV color for starburst galaxies. The
  plot style is the same as in Figure~\ref{figure.fuvnuv_normal}.}
\label{figure.fuvnuv_sb}
\end{center}
\end{figure}

In addition to structures visible in each band, we looked for spectral
structure in the UV halos. In regions where the halo is detected in
all five filters we can construct SEDs, but most galaxies in the
sample do not have enough data. Thus, we adopt the FUV$-$NUV color as
a proxy.  Figures~\ref{figure.fuvnuv_normal} and
\ref{figure.fuvnuv_sb} show the FUV$-$NUV color in regions where the
flux is detected at 2$\sigma$ above background in each filter for
normal and starburst galaxies, respectively.  In these images, the
smoothing kernel is usually larger than in
Figures~\ref{figure.gallery_normal} and
\ref{figure.gallery_starburst}. The colorbar in the plots is such that
blue regions correspond to a blue spectrum (small FUV-NUV) and red
regions correspond to a red spectrum (large FUV-NUV).  However, values
at the edges are not reliable as they are sensitive to the background
subtraction (e.g., the red boundaries on the halo of NGC~55 in
Figure~\ref{figure.gallery_normal}).

There is no obvious structure in the FUV$-$NUV color maps for most
normal galaxies. There appears to be a tendency for the Sc-type
galaxies (NGC~5907, NGC~4522, NGC~4330, and NGC~3044) to have bluer
halos than the rest (which are Sb, except for NGC~55), but the
strength of this connection cannot be assessed with the current
sample.  This is especially true since NGC~4522 and NGC~4330 are being
stripped of their gas by ram pressure from the Virgo intracluster
medium.  For galaxies where the $S/N$ is reasonably high across the
disk (NGC~55, NGC~891, NGC~4217, and NGC~4330), the color does not
change much across with projected galactocentric radius. However, in
the starburst galaxies the filaments and regions associated with
galactic winds are bluer than the rest of the halo. The underlying
FUV-NUV color differs between galaxies, with NGC~3628 and NGC~4388
having generally redder halos; both of these galaxies are classified
as Sb type, whereas the others are Sc or Sd. Combined with the normal
galaxies, this suggests that bluer galaxies tend to have bluer halos,
but the sample size is small.
In M82 the FUV-NUV color tends
to decrease with height, but the filamentary structures within the
wind that are visible in both the FUV and NUV images do not appear in
the FUV-NUV color map. In other words, small UV structures in the wind
appear to have the same color as the rest of it. It is also of
interest that the bluer regions in the halo of NGC~4631 are asymmetric
across the midplane, although both sides are blue in the central
region where the wind occurs.

\subsection{Multiwavelength Comparison}

We compare the diffuse UV halos to diffuse extraplanar
light in other bands, as well as the disk emission in the same
bands. These include H$\alpha$, soft X-rays, 160\,$\mu$m emission,
radio continuum (at 1.4 or 4.8\,GHz) and 21\,cm emission from neutral
hydrogen.  The H$\alpha$ traces the diffuse ionized gas in and above
the disk, and the presence of extraplanar diffuse ionized gas (eDIG)
indicates disk porosity and ongoing star formation (it may also
produce UV emission nebulae through ionized helium). We use H$\alpha$
images from several atlases
\citep{lehnert96,rand96,collins00,rossa03b}. Diffuse soft X-rays trace
hot gas, and X-ray halos seen near the disk are primarily from
outflows connected to star formation \citep{li13}. Within the disk,
some of the X-rays also come from X-ray binaries. We use X-ray data
from the \textit{Chandra} and \textit{XMM-Newton} archives, and the
presence of X-ray halos is determined from the literature
\citep{strickland04,tullmann06,li13}. The 160\,$\mu$m images trace
thermal emission from dust, and we obtain them from the
\textit{Herschel} archive. We use 160$\mu$m because it is near the
expected peak of the modified blackbody curve for thermal emission,
and the spatial resolution at longer wavelengths is considerably
worse. The 160$\mu$m images are dominated by the disk and the PSF is
about 6\,arcsec, so we adopt the same approach to clean the overlapping
Airy patterns and isolate extraplanar emission (except in
NGC~4522). We use the PSF from \citet{aniano11}, and in some cases the
galaxy box is smaller than for the UV.  We also refer to
\citet{mccormick13} regarding extraplanar polycyclic aromatic
hydrocarbon (PAH) emission and \citet{howk97} and \citet{howk99} for
the presence of extraplanar dust seen in absorption as filaments in
the optical band. The radio continuum emission comes from cosmic rays
produced in supernova remnants, which diffuse out of the disk. We
primarily refer to data taken as part of the CHANG-ES program
\citep{wiegert15}, but some images come from prior atlases
\citep[especially][]{condon87}. Finally, we use or refer to high
resolution 21\,cm maps from a variety of papers
\citep{weliachew78,haynes79,irwin87,irwin94,yun94,lee97,kenney01,walter04,oosterloo07,chung09,zschaechner12,westmeier13,yim14,allaert15,lucero15,vollmer16}. These
maps often show filaments of cold gas or warped disks.

\begin{figure*}
\begin{center}
\includegraphics[width=0.95\textwidth]{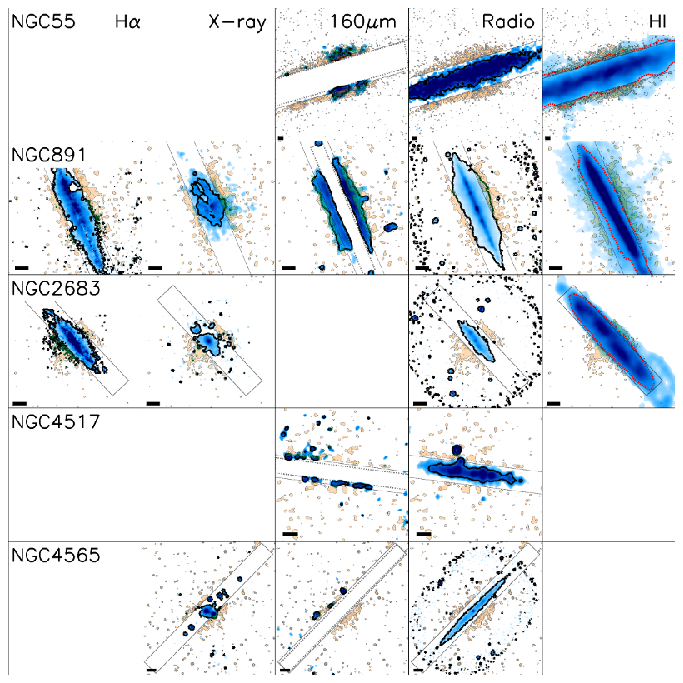}
\caption{\scriptsize A comparison of the UV halos to diffuse emission
  in other bands for normal galaxies. Empty panels indicate where we
  have no image file, but in some cases we can compare the morphology
  from published images (see Table~\ref{table.obs}). The orange maps
  show the FUV (or \uvwtwo{}) contours from Figure~\ref{figure.gallery_normal} while
  the blue maps show images at the other wavebands. The thick black contour is the
  3$\sigma$ contour above background for each of these images. The
  thin black box shows our galaxy region for PSF-wing
  contamination. To emphasize the extraplanar emission at 160\,$\mu$m
  we also corrected these images for PSF-wing contamination, and the
  dotted lines show the galaxy regions used. In some images the map
  edges are visible as regions of heightened noise, and the contours
  there should be ignored.}
\label{figure.multiwave_normal}
\end{center}
\end{figure*}
\setcounter{figure}{7}

\begin{figure*}
\begin{center}
\includegraphics[width=0.95\textwidth]{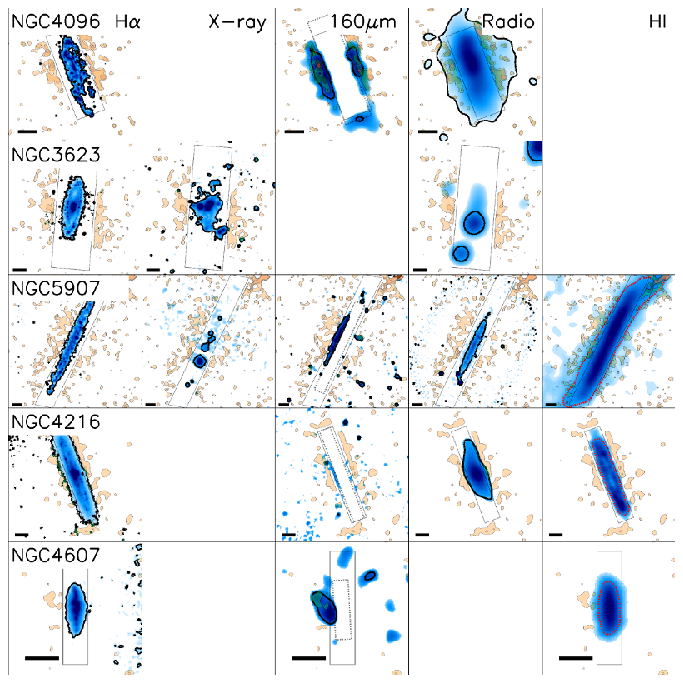}
\caption{\scriptsize continued}
\end{center}
\end{figure*}
\setcounter{figure}{7}

\begin{figure*}
\begin{center}
\includegraphics[width=0.95\textwidth]{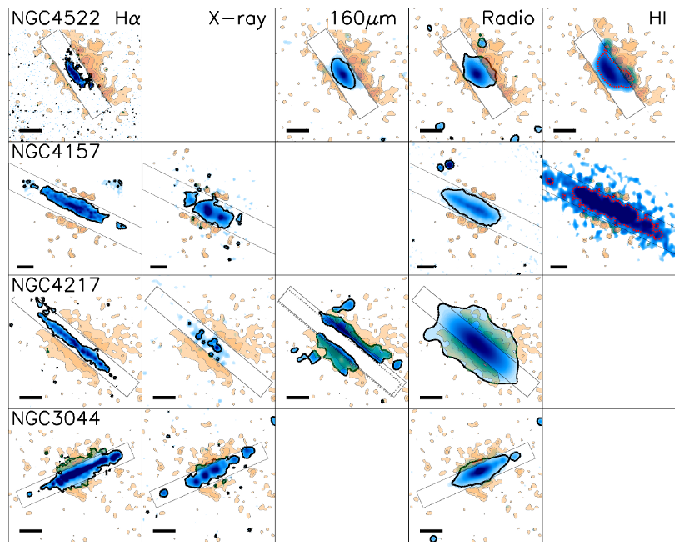}
\caption{\scriptsize continued}
\end{center}
\end{figure*}

H$\alpha$, soft X-rays, thermal dust emission, and the radio continuum
are associated with ongoing star formation, whereas \ion{H}{1} need
not be.  Figures~\ref{figure.multiwave_normal} and
\ref{figure.multiwave_sb} show the UV contours from above overlaid on
H$\alpha$, X-ray, 160\,$\mu$m, radio continuum, and 21\,cm images for
normal and starburst galaxies, respectively. In some cases there is no
publicly available data, but we can usually rely on images in the
literature (this is especially true for 21\,cm data). Galaxies are not
included in these figures if they do not have publicly available FITS
files in at least two bands. As with the UV images, the $S/N$ is
variable among the galaxies and bands, which limits the power of this
comparison.

\paragraph{Extraplanar Diffuse Emission: Frequency}
About 40\% of non-starburst galaxies have eDIG \citep{rossa03a}, and a
similar number of galaxies in our sample with data from these bands
(Table~\ref{table.derived}) have eDIG (38\%), radio (45\%), or soft
X-ray halos (40\%). However, 67\% of galaxies with 160\,$\mu$m images
have extraplanar emission (sometimes at lower heights than the UV
halos; Figure~\ref{figure.multiwave_normal}).  The sample overlap with
\citet{mccormick13} is not large enough to determine the fraction with
extraplanar PAH emission. In contrast, UV halos are seen around every
normal galaxy. Several of the galaxies in our sample were also
investigated by \citet{howk97} and \citet{howk99}, who used unsharp
mask images in the optical to identify high latitude dust
absorption. These include NGC~891, NGC~4517, NGC~4565, NGC~5907,
NGC~4217, and NGC~4157, of which NGC~891 and NGC~4217 have high
latitude dust. The starburst galaxies all have eDIG, soft X-ray halos,
extraplanar thermal dust emission (except for NGC~3628), and radio halos.
\ion{H}{1} emission is also seen around all galaxies in the sample
with high resolution maps, although in some cases the beamsize makes
it unclear how much of the emission is truly extraplanar.

\paragraph{Extraplanar Diffuse Emission: Morphology}
The extraplanar diffuse emission in normal galaxies shows some of the same
features as the UV halo morphology, although sometimes it is only detected at
lower heights where we cannot measure the UV halo morphology
directly. The eDIG tends to have a
similar morphology as the UV in terms of filaments, concentration, and
bilateral symmetry, but it is usually not detected at larger heights
where the UV halo remains bright. The soft X-ray halos correspond
reasonably well to the UV halo structure in NGC~891, NGC~2683, and
NGC~4217, but not in NGC~5170. There are fewer normal galaxies with
X-ray data than with H$\alpha$ images, and only four with X-ray halos,
so it is unclear how well the morphology matches the UV. 
The extraplanar 160\,$\mu$m emission coincides with the UV,
but it is not always as extensive. NGC~55, NGC~891,
and NGC~4096 have the most convincing 160\,$\mu$m halos, but their
prominence does not appear to be connected to that in H$\alpha$ or the
radio continuum.  Likewise, the prominence of the radio
halos seems unrelated to the prominence of the UV halo or extraplanar
emission in other bands.  The only normal galaxies in
\citet{mccormick13} with extraplanar PAH emission are NGC~55 and
NGC~891. In the former case, the PAHs do not follow the UV halo,
whereas in the latter they do. NGC~891 and NGC~4217 have high latitude
dust seen by \citet{howk99}. Although these two galaxies do have
brighter UV halos than others in the \citet{howk99} sample, the UV
halos in the four galaxies without high latitude dust absorption have
about the same vertical extent and radial concentration.

Starburst UV halos have filamentary structure that is also seen at other wavelengths. This is
especially true in H$\alpha$, although a similar pattern is seen in
X-ray and radio halos. There are some disagreements in extraplanar
morphology among these bands (such as in M82, NGC~4631, or NGC~4388),
but there is a clear connection between the UV filaments and those
seen at other wavelengths. The same is true for the extraplanar dust
emission, although not every filament is associated with 160\,$\mu$m
emission. However, unlike at other wavelengths the UV halos span the
disks of the starburst galaxies, and this non-filamentary structure is
not present in H$\alpha$, X-rays, thermal dust emission, or radio
continuum. In general the PAHs do not follow the UV halos \citep{mccormick13}.
Some of the same filaments are visible in PAHs, but the coincidence of UV
and PAH filaments is worse than at other wavelengths, and the diffuse
components above the disk have different shapes.

The connection between the 21-cm morphology and the UV halos is
unclear.  We note that the sensitivities and resolutions of the
\ion{H}{1} images in Figure~\ref{figure.multiwave_sb} differ, so the
$N_{\text{H}} = 5\times 10^{20}$\,cm$^{-2}$ contour (assuming
optically thin hydrogen) is shown as a shorthand for the transition
from disk to halo gas. The \ion{H}{1} and UV morphology are similar
for the galaxies undergoing ram-pressure stripping (NGC~134, NGC~4522,
NGC~4330, and NGC~4388). The galaxies in the Virgo cluster also tend
to have less extensive extraplanar \ion{H}{1}, but there is no clear
difference between the UV halos of the Virgo galaxies (NGC~4313,
NGC~4216, NGC~4330, NGC~4607, NGC~4522, NGC~4217, NGC~4388, and
NGC~5775) and the others. There is also no apparent connection between
the UV halo morphology and \ion{H}{1} warps, or between the vertical
concentration in the \ion{H}{1} total column maps and the diffuse UV
images. 

\begin{figure*}
\begin{center}
\includegraphics[width=0.95\textwidth]{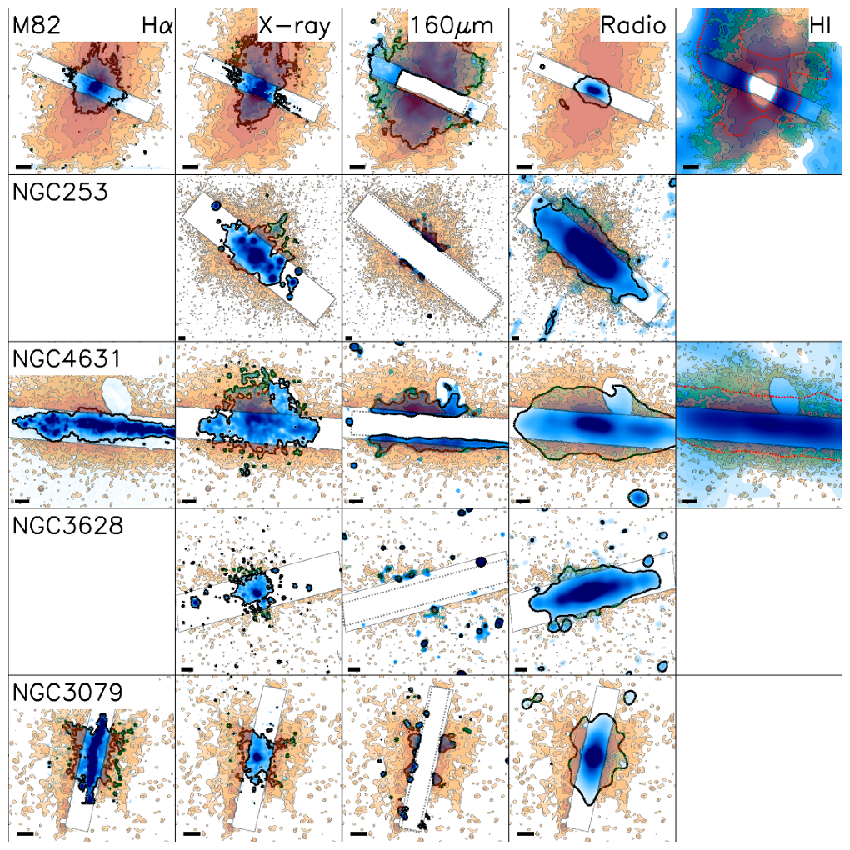}
\caption{\scriptsize A comparison of the UV halos to diffuse emission
  in other bands for starburst galaxies. The plot style is the same as
in Figure~\ref{figure.multiwave_normal}.}
\label{figure.multiwave_sb}
\end{center}
\end{figure*}
\setcounter{figure}{8}

\begin{figure*}
\begin{center}
\includegraphics[width=0.95\textwidth]{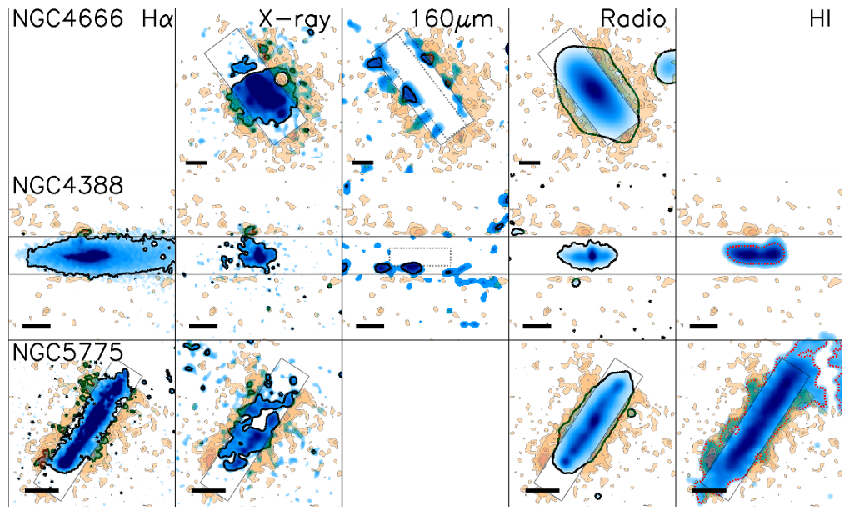}
\caption{\scriptsize continued}
\end{center}
\end{figure*}

The UV and \ion{H}{1} filaments do coincide near the disk in the starburst
galaxies NGC~253, NGC~3628, NGC~4666, NGC~5775, and NGC~3079 (not all
shown in Figure~\ref{figure.multiwave_sb}). The UV
halo is also brighter and more vertically extended on the side of the
galaxy with the large \ion{H}{1} filament in NGC~891, and a similar
phenomenon is seen in NGC~3044.  However, the large-scale hydrogen
filaments seen around galaxies in our sample (such as NGC~55,
NGC~3628, NGC~4565, NGC~4631, or M82) are not visible in the UV. 

\paragraph{Disk Emission}
There is good agreement in most normal galaxies between the projected
radial extent (along the disk) of the bright H$\alpha$ disk emission
and the UV halo above it, which mostly explains the difference in UV
halo concentration among normal galaxies.  The same is true for soft
X-rays, radio continuum, and thermal emission from dust in the disk,
except for NGC~4565 and NGC~4522
(Figure~\ref{figure.multiwave_normal}).  In the starburst galaxies,
the disk emission is less connected to the UV halo and is instead
brightest in the regions where the galactic winds occur. 

\subsection{Summary}

UV halos around luminous spiral galaxies are
ubiquitous.  In normal galaxies, the UV halos tend to form a thick
disk. Their presence is independent of diffuse extraplanar material
seen in other wavebands, but in those cases where a diffuse
extraplanar layer does exist, it has a similar morphology to the UV
halo and occurs in about the same place. This suggests that the UV
halos arise from some gaseous medium, which is supported by the
asymmetric halos in NGC~4522 and NGC~4330 that match the morphology of
the stripped gas, but not the morphology of the stellar disk.  The
FUV$-$NUV colors differ between galaxies, but there are no clear
structures in the FUV$-$NUV colors, except perhaps in stripped
galaxies. The UV halos do appear to be connected to star formation in
the disk, as they are brightest over the parts of the disk with bright
H$\alpha$, X-ray, or radio emission, but not in a way that requires
ionizing photons or winds to escape the disk. 

UV halos around starburst galaxies are more extended and have 
filamentary structures coincident with the galactic winds. Apart from
these filaments, there is little agreement between the extraplanar or
disk emission in other wavebands. The winds are visible in the FUV$-$NUV
maps as bluer regions in the UV halo, and may be
superimposed on a disk structure like that in the normal galaxies, but
the origin of the color difference is not clear.

\section{Ultraviolet Halo Fluxes}
\label{section.fluxes}

In the prior section we focused on qualitative morphological trends.
Here we compare the average halo fluxes and FUV$-$NUV colors, as
well as the flux as a function of height above the disk, to various
metrics for each galaxy. It is especially important to understand the
connection between the structure visible in the UV maps and the
average properties because it will not be possible to observe this
structure with reasonable exposure times for more distant galaxies.

We measured halo UV fluxes in long boxes parallel to the midplane of
the galaxy, as in \citetalias{hk14}. The length of each box is equal
to the length of the region containing the galaxy image, and the width
of each box is a projected 2\,kpc at the distance of the galaxy. We
also reference the height of the box to the projected midplane of the
galaxy. For galaxies with lower inclinations, such as NGC~4666, the
projected heights are too large. The halo fluxes are all measured
starting from a projected height of 2-5\,kpc from the midplane in our
sample. This conservative threshold excludes galaxy disk light, but
likely misses halo light at lower latitudes. The galaxy fluxes
measured in the boxes used for PSF-wing correction are reported in
Table~\ref{table.galaxy_fluxes}.

The halo fluxes are corrected for PSF-wing contamination, and the
\uvwtwo{} fluxes are corrected for the red leak using a galaxy-type
correction from \citetalias{hk14} to isolate the true UV flux
(shortward of 3000\AA). We used a correction factor of 0.93, except
for M82 (0.95) and NGC~3623 (0.85). The \uvwone{} filter has a more
severe red leak for which this correction is unreliable, so we report
the uncorrected fluxes. The total halo fluxes, amount of PSF-wing
contamination, and \uvwtwo{} correction factors are listed in
Table~\ref{table.fluxes}.

In the remainder of this section, we assess the reliability of the
measured fluxes, measure scale heights for the UV halos, and examine
the change in color as a function of height. We then search for
correlations between the average UV halo fluxes and galactic
properties, and compare our results to \citetalias{hk14}.

\subsection{Reliability of Halo Fluxes}

The similarity in the halo morphology between different UV filters
suggests that our PSF-wing subtraction is accurate, but it is worth
considering the reliability of the measured fluxes when the PSF-wing
contamination can be a large fraction of the total flux. We performed
several tests, and conclude that the fluxes are indeed reliable in the
sense that the systematic error from PSF-wing subtraction is smaller
than the statistical error.

\begin{deluxetable*}{lcccccc}
\tablenum{4}
\tabletypesize{\scriptsize}
\tablecaption{Galaxy UV Fluxes}
\tablewidth{0pt}
\tablehead{
\colhead{Name}	& \colhead{FUV}	& \colhead{NUV} & \colhead{\uvwtwo{}} & \colhead{\uvmtwo{}} & \colhead{\uvwone{}} & \colhead{\uvwtwo{} Factor}\\
\colhead{(1)} & \colhead{(2)} & \colhead{(3)} & \colhead{(4)} & \colhead{(5)} & \colhead{(6)} & \colhead{(7)} 
}
\startdata
\cutinhead{Starburst Galaxies}
NGC0253	& $105.20\pm0.01$	& $200.78\pm0.02$	& 					& 					& 					& 0.93\\
M82		& $3.67\pm0.04$		& $17.30\pm0.08$	& $14.47\pm0.09$	& $13.49\pm0.07$	& $37.96\pm0.04$	& 0.95\\
NGC4631	& $92.80\pm0.06$	& $132.3\pm0.1$		& $108.4\pm0.2$		& $121.3\pm0.3$		& $132.6\pm0.4$		& 0.93\\
NGC3628	& $5.45\pm0.02$		& $14.6\pm0.1$		& $11.6\pm0.1$		& $11.9\pm0.2$		& $28.3\pm0.9$		& 0.93\\
NGC4666	& $5.69\pm0.04$		& $10.20\pm0.04$	& $8.31\pm0.07$		& $8.20\pm0.09$		& $13.95\pm0.08$	& 0.93\\
NGC3079	& $10.39\pm0.01$	& $16.29\pm0.02$	& $13.96\pm0.02$	& $16.12\pm0.04$	& $20.9\pm0.1$		& 0.93\\
NGC5775	& $2.47\pm0.01$		& $4.48\pm0.03$		& $4.38\pm0.01$		& $4.13\pm0.02$		& $7.36\pm0.03$		& 0.93\\
NGC4388	& $3.16\pm0.01$		& $6.50\pm0.01$		& $5.68\pm0.01$		& $6.36\pm0.01$		& $9.87\pm0.02$		& 0.93\\
\cutinhead{Normal Galaxies}																									
NGC0055	& $294.56\pm0.01$	& $441.60\pm0.03$	& 					& 					& 					& 0.93\\
NGC0891	& $2.85\pm0.06$		& $5.7\pm0.2$		& $5.5\pm0.2$		& $4.6\pm0.3$		& $10.77\pm0.05$	& 0.93\\
NGC2683	& $6.36\pm0.02$		& $12.35\pm0.04$	& $10.27\pm0.04$	& $10.23\pm0.09$	& $19.9\pm0.3$		& 0.93\\
NGC4517	& $6.39\pm0.02$		& $10.27\pm0.03$	& 					& 					& 					& 0.93\\
NGC4565	& 					& $15.45\pm0.01$	& 					& 					& 					& 0.93\\
NGC4096	& $7.26\pm0.03$		& $12.80\pm0.04$	& $10.39\pm0.03$	& $11.17\pm0.03$	& 					& 0.93\\
NGC4313	& $0.24\pm0.01$		& $1.00\pm0.01$		& 					& 					& 					& 0.93\\
NGC3623	& $3.25\pm0.04$		& $9.58\pm0.05$		& $7.60\pm0.03$		& $8.04\pm0.02$		& $20.66\pm0.05$	& 0.85\\
NGC5907	& $7.99\pm0.03$		& $12.96\pm0.05$	& $10.99\pm0.03$	& $11.58\pm0.06$	& $16.75\pm0.04$	& 0.93\\
NGC4216	& $1.42\pm0.02$		& $3.70\pm0.02$		& 					& 					& 					& 0.93\\
NGC4607	& $0.22\pm0.01$		& $0.60\pm0.01$		& 					& 					& 					& 0.93\\
NGC4522	& $2.57\pm0.01$		& $4.03\pm0.02$		& 					& 					& 					& 0.93\\
NGC0134	& 					& 					& $10.28\pm0.05$	& $10.46\pm0.07$	& $17.4\pm0.2$		& 0.93\\
NGC4157	& $3.13\pm0.02$		& $9.60\pm0.04$		& 					& 					& 					& 0.93\\
ESO358	& $1.62\pm0.01$		& $3.30\pm0.02$		& 					& 					& 					& 0.93\\
NGC4217	& $0.74\pm0.03$		& $1.50\pm0.03$		& 					& 					& 					& 0.93\\
NGC4330	& $1.41\pm0.01$		& $2.27\pm0.01$		& $1.87\pm0.02$		& $1.99\pm0.02$		& $2.64\pm0.03$		& 0.93\\
NGC3044	& $5.77\pm0.02$		& $9.30\pm0.02$		& $6.98\pm0.04$		& $7.74\pm0.03$		& $9.56\pm0.03$		& 0.93\\
NGC5170	& $2.40\pm0.07$		& $5.06\pm0.04$		& 					& 					& 					& 0.93
\enddata
\tablecomments{\label{table.galaxy_fluxes} The galaxy fluxes are measured
  in boxes as described in Section~\ref{section.fluxes}. The
  \uvwtwo{} scale factor is based on the galaxy type as described in \citetalias{hk14}. }
\end{deluxetable*}

First, the PSF-wing contamination is not very sensitive to the
best-fit exponent because most of the PSF-wing contamination comes
from the region where the profile is measured rather than
extrapolated. If we extrapolate the PSFs with an exponent of $x=2$ in
each filter, which is unacceptably shallow (Table~\ref{table.psf}) and
inconsistent with the background, the fractional increase in the 
PSF-wing contamination is only about 2\%, except in
the \uvwone{} filter where the change is about 8\%.

Second, the NUV filter covers approximately the same wavelength range
as the combined \uvmtwo{} and \uvwone{} filters. Since the PSF-wing
contamination is smaller in the NUV, comparing the NUV fluxes to the
average of the \uvmtwo{} and \uvwone{} can tell us if the PSF-wing
contamination is obviously too high or low. For galaxies with fluxes
in each filter we measured the flux with and without accounting for
PSF-wing contamination, and averaged the \uvmtwo{} and \uvwone{}
values. Without the PSF-wing correction, the mean UVOT/\Galex{} flux
ratio is about 1.6 across our sample, whereas after correction it is
about 1.05. We cannot exactly synthesize the NUV filter from the UVOT
filters, so a ratio of 1.0 is not expected and this comparison cannot tell us if
there are small systematic uncertainties. Nonetheless, this exercise
suggests that the reported fluxes are close to the true flux.

Finally, there are \textit{HST} observations of the halo of M82 in
several UV bands (using the Wide Field Camera 3 UVIS detector),
including the F225W filter that overlaps with the \uvwtwo{}, NUV,
\uvmtwo{}, and \uvwone{} filters. The sub-arcsecond angular resolution and
small field of view mean that PSF-wing contamination is
negligible. The wind is clearly visible in the \textit{HST} image, and the flux measured in this
region is consistent with that measured with \Swift{} and \Galex{}
after PSF-wing subtraction but not before. For example, in one region
(a circle of radius 15\,arcsec centered on $\alpha =$09:55:52.22,
$\delta = +$69:40:02.39), we measure an AB magnitude of
$m_{\text{F225W}} = 16.91\pm0.04$\,mag in the \textit{HST} image,
$m_{\text{uvm2}} = 16.80\pm0.03$\,mag in the \uvmtwo{} image before
PSF-wing correction, and $m_{\text{uvm2}} = 16.92\pm0.03$\,mag
after. Similar values are found in the other filters, indicating that
the PSF-wing subtraction recovers the true flux to within the
statistical uncertainty. 

\begin{deluxetable*}{lccccccccccc}
\tablenum{5}
\tabletypesize{\scriptsize}
\tablecaption{Halo UV Fluxes}
\tablewidth{0pt}
\tablehead{
\colhead{}      & \multicolumn{5}{c}{Halo Flux (mJy)} & \multicolumn{5}{c}{PSF-Wing Contamination (mJy)} & \colhead{\uvwtwo{} Factor} \\
\colhead{Name}	& \colhead{FUV}	& \colhead{NUV} & \colhead{\uvwtwo{}} & \colhead{\uvmtwo{}} & \colhead{\uvwone{}} &	\colhead{FUV}	& \colhead{NUV} & \colhead{\uvwtwo{}} & \colhead{\uvmtwo{}} & \colhead{\uvwone{}} &	\\
\colhead{(1)} & \colhead{(2)} & \colhead{(3)} & \colhead{(4)} & \colhead{(5)} & \colhead{(6)} & \colhead{(7)} & \colhead{(8)} & \colhead{(9)} & \colhead{(10)} & \colhead{(11)} & \colhead{(12)} 
}
\startdata
\cutinhead{Starburst Galaxies}
NGC0253	& $5.43\pm0.09$	& $17.2\pm0.2$	& 				& 				& 				& 4.89	& 0.60	& 		& 		& 		& 0.93\\
M82		& $7.0\pm0.1$	& $15.7\pm0.2$	& $14.4\pm0.2$	& $13.3\pm0.2$	& $25.9\pm0.1$	& 0.29	& 0.53	& 0.69	& 0.54	& 2.94	& 0.95\\
NGC4631	& $2.74\pm0.07$	& $5.3\pm0.2$	& $5.0\pm0.3$	& $5.0\pm0.4$	& $6.20	\pm0.5$	& 5.75	& 0.99	& 3.16	& 3.01	& 6.39	& 0.93\\
NGC3628	& $0.46\pm0.01$	& $1.92\pm0.09$	& $1.4\pm0.1$	& $1.2\pm0.1$	& $3.3\pm0.4$	& 0.20	& 0.11	& 0.21	& 0.19	& 0.80	& 0.93\\
NGC4666	& $0.23\pm0.02$	& $0.51\pm0.02$	& $0.49\pm0.03$	& $0.36\pm0.04$	& $0.48\pm0.04$	& 0.16	& 0.06	& 0.13	& 0.11	& 0.39	& 0.93\\
NGC3079	& $0.42\pm0.01$	& $0.65\pm0.02$	& $0.65\pm0.05$	& $0.63\pm0.04$	& $0.5\pm0.1$	& 0.53	& 0.23	& 0.43	& 0.41	& 1.15	& 0.93\\
NGC5775	& $0.23\pm0.01$	& $0.39\pm0.02$	& $0.49\pm0.01$	& $0.40\pm0.01$	& $0.71\pm0.03$	& 0.13	& 0.11	& 0.17	& 0.14	& 0.45	& 0.93\\
NGC4388	& $0.12\pm0.01$	& $0.37\pm0.01$	& $0.25\pm0.01$	& $0.29\pm0.01$	& $0.90\pm0.02$	& 0.14	& 0.08	& 0.16	& 0.14	& 0.49	& 0.93\\
\cutinhead{Normal Galaxies}																									
NGC0055	& $1.0\pm0.2$	& $12.6\pm0.8$	& 				& 				& 				& 13.99	& 0.51	& 		& 		& 		& 0.93\\
NGC0891	& $0.58\pm0.05$	& $0.8\pm0.2$	& $0.8\pm0.1$	& $1.0\pm0.3$	& $1.3\pm0.2$	& 0.15	& 0.05	& 0.15	& 0.10	& 0.47	& 0.93\\
NGC2683	& $0.30\pm0.02$	& $0.88\pm0.04$	& $0.84\pm0.04$	& $0.5\pm0.1$	& $1.6\pm0.4$	& 0.29	& 0.11	& 0.28	& 0.22	& 0.94	& 0.93\\
NGC4517	& $0.23\pm0.02$	& $0.9\pm0.4$	& 				& 				& 				& 0.33	& 0.12	& 		& 		& 		& 0.93\\
NGC4565	& 				& $1.08\pm0.09$	& 				& 				& 				& 		& 0.17	& 		& 		& 		& 0.93\\
NGC4096	& $0.10\pm0.02$	& $0.48\pm0.04$	& $0.54\pm0.03$	& $0.41\pm0.02$	& 				& 0.34	& 0.18	& 0.31	& 0.27	& 		& 0.93\\
NGC4313	& $0.05\pm0.01$	& $0.12\pm0.02$	& 				& 				& 				& 0.01	& 0.03	& 		& 		& 		& 0.93\\
NGC3623	& $0.02\pm0.02$	& $0.21\pm0.03$	& $0.19\pm0.02$	& $0.17\pm0.01$	& $0.47\pm0.03$	& 0.06	& 0.01	& 0.07	& 0.06	& 0.37	& 0.85\\
NGC5907	& $0.22\pm0.02$	& $0.43\pm0.04$	& $0.39\pm0.02$	& $0.43\pm0.04$	& $0.42\pm0.03$	& 0.39	& 0.10	& 0.29	& 0.25	& 0.77	& 0.93\\
NGC4216	& $0.08\pm0.01$	& $0.34\pm0.02$	& 				& 				& 				& 0.07	& 0.08	& 		& 		& 		& 0.93\\
NGC4607	& $0.02\pm0.01$	& $0.07\pm0.01$	& 				& 				& 				& 0.01	& 0.03	& 		& 		& 		& 0.93\\
NGC4522	& $0.15\pm0.01$	& $0.25\pm0.02$	& 				& 				& 				& 0.11	& 0.09	& 		& 		& 		& 0.93\\
NGC0134	&  				& 				& $0.44\pm0.03$	& $0.24\pm0.05$	& $0.6\pm0.1$	& 		& 		&  0.26	& 0.21	& 0.75	& 0.93\\
NGC4157	& $0.15\pm0.02$	& $0.31\pm0.01$	& 				& 				& 				& 0.13	& 0.08	& 		& 		& 		& 0.93\\
ESO358	& $0.06\pm0.01$	& $0.19\pm0.02$	& 				& 				& 				& 0.06	& 0.03	& 		& 		& 		& 0.93\\
NGC4217	& $0.17\pm0.03$	& $0.50\pm0.03$	& 				& 				& 				& 0.04	& 0.06	& 		& 		& 		& 0.93\\
NGC4330	& $0.11\pm0.01$	& $0.17\pm0.01$	& $0.20\pm0.03$	& $0.15\pm0.02$	& $0.26\pm0.05$	& 0.08	& 0.16	& 0.12	& 0.11	& 0.26	& 0.93\\
NGC3044	& $0.11\pm0.02$	& $0.13\pm0.01$	& $0.09\pm0.04$	& $0.10\pm0.03$	& $0.02\pm0.05$	& 0.31	& 0.32	& 0.29	& 0.28	& 0.69	& 0.93\\
NGC5170	& $0.04\pm0.05$	& $0.67\pm0.05$	& 				& 				& 				& 0.08	& 0.09	& 		& 		& 		& 0.93
\enddata
\tablecomments{\label{table.fluxes} The halo fluxes are
  measured as described in Section~\ref{section.fluxes}. The
  \uvwtwo{} scale factor is based on the galaxy type as described in \citetalias{hk14}. The 
  total flux measured on the CCD is the sum of the inferred astrophysical flux (``halo flux'') and the
  PSF-wing contamination.}
\end{deluxetable*}

\subsection{Scale Heights}

UV halos appear to have a morphology like a thick disk, so we measured
scale heights ($h$) based on fitting a function of the form $F(z) = A
e^{-|z|/h}$, where $F$ is the flux, $A$ is a normalization, and $z$ is
the projected height above the midplane. We used the fluxes measured
in long boxes because the $S/N$ is only high enough in a few galaxies
to measure scale heights at different radii. 
To see how the aggregate behaves, we fit scale heights at several galactocentric
radii in the stacked quadrant images. Figure~\ref{figure.quadrants} shows the
scale height as a function of galactocentric radius, along with the 1$\sigma$
error bars, for the normal and starburst galaxies. In the starburst galaxies
the source mask near $R=0$ for NGC~4631 is taken into account.
In the normal galaxy stack the scale height within $R_{25}$ is
consistent with being constant, with a possible increase with larger radii
that may be due to a more complex background that results from the stacking.
The starburst stack shows that the scale height changes with galactocentric
radius, which can be explained by filaments. 
The scale heights can also differ between filters, as is the
case for M82 and NGC~253, but in most galaxies the scale heights
measured in different filters agree within the uncertainty. Here we
report the value of $h$ measured from the average radial profile and
combining each filter (we allowed $A$ to vary
between filters, but fit for a common $h$). Table~\ref{table.derived}
contains the best-fit $h$ and the associated reduced $\chi^2$.

Scale heights range from $h=1-8$\,kpc. The mean value of $h$ for 
starburst galaxies is $\bar{h}=3.6$\,kpc ($3.0$\,kpc when excluding
NGC~4666, which has a lower inclination), and it is $\bar{h}=3.5$\,kpc
for normal galaxies. There is no dependence on H$\alpha$
luminosity, the presence of eDIG, Hubble type, UV halo luminosity, or
the $S/N$ in the image. Galaxies with stripped gas (NGC~4522, NGC~134,
and NGC~4330) have small scale heights ($h<2$\,kpc), but NGC~891 and
NGC~55 have similar $h$ values even without stripped gas. Likewise, there
is no connection between $h$ and the fraction of UV light from
filaments. The galactic or halo properties that seem to be connected
to other morphological indicators of UV halos do not predict the scale
height.

In several galaxies, $h$ varies between filters.
For example, in NGC~253 the joint fit to all filters results in $h=2.2$
but with a reduced $\chi^2 = 38$. The FUV profile is marginally fit
for $h=1.8$\,kpc (reduced $\chi^2_{\nu} = 2$), but the NUV profile cannot be
fit with a single exponential profile ($h=2.6$\,kpc with reduced
$\chi^2 = 43$). A double exponential profile is a good fit with an
inner component ($h\approx 1$\,kpc) and an outer component ($h \approx
3$\,kpc), so it is possible that the NUV light has a compact component
and a component similar to the FUV
halo. Figure~\ref{figure.gallery_starburst} shows that the NUV light
follows the disk much better than the FUV light, suggesting a possible
difference in physical origin. A similar difference is found in M82,
where the joint fit produces $h=2.7$\,kpc ($\chi^2_{\nu} = 32$) and
the best-fit values of $h$ in each filter are: $h=4.9$, 3.0, 2.9, 3.2,
and 2.5\,kpc for the FUV, \uvwtwo{}, \uvmtwo{}, NUV, and \uvwone{}
filters, respectively (with corresponding $\chi^2_{\nu} = 1.7$, 4.4,
0.8, 2.9, and 12). As in NGC~253, the FUV halo is more extended, and the
redder filters cannot be adequately described by single exponential
models. On the other hand, in NGC~55 (joint $h=1.1$\,kpc
and $\chi^2_{\nu} = 4.2$) the FUV halo has a smaller scale height
than the NUV ($h \approx 0.5$\,kpc compared to 1.3\,kpc). It is
tempting to ascribe the difference in behavior to the presence of
galactic winds, and indeed NGC~3079, NGC~4631, and NGC~4666 do have
shallower FUV than NUV profiles, but the $S/N$ is generally too low in
the normal galaxies to determine if this is a clear difference between
them.

We also searched for different scale heights across the midplane. In
many cases, one filter has a different profile across the midplane
while the others do not, but these can usually be explained by a small
background gradient across the region of the galaxy (e.g., due to
Galactic cirrus, which differs in strength between filters). The
exception is M82, where the FUV markedly differs across the midplane
in a way that other filters do not. The most likely explanation is
that the wind is brightest in the FUV, as seen in
Figure~\ref{figure.fuvnuv_sb}.  Most galaxies have approximately
symmetric flux profiles, despite differences in the visible morphology
across the midplane (e.g., Figure~\ref{figure.gallery_starburst}).
The galaxies where there is a notable difference in multiple filters
include NGC~4666, NGC~4522, NGC~4330, and NGC~134. The latter three
galaxies have an obvious asymmetry due to stripped gas
(Figure~\ref{figure.gallery_normal}). NGC~4666 is the least inclined
galaxy in the sample, and the asymmetry in scale heights may pertain
to inclination and projection effects, as mentioned in
\citetalias{hk14}.

\subsection{FUV-NUV Color with Height}

We measured the correlation between the FUV$-$NUV color in each flux
measurement bin and the projected height
of the bin for each galaxy with data in both filters. Most galaxies
have no significant correlation (for a threshold of $p=0.05$), and the
exceptions are all starbursts. In NGC~253, M82, NGC~4631, NGC~3628,
NGC~3079, and NGC~4388, the FUV$-$NUV color decreases with
height. This is consistent with the FUV$-$NUV color maps
(Figure~\ref{figure.fuvnuv_sb}) and suggests that the UV halos
comprise two components. 

\begin{figure}
\begin{center}
\includegraphics[width=0.5\textwidth]{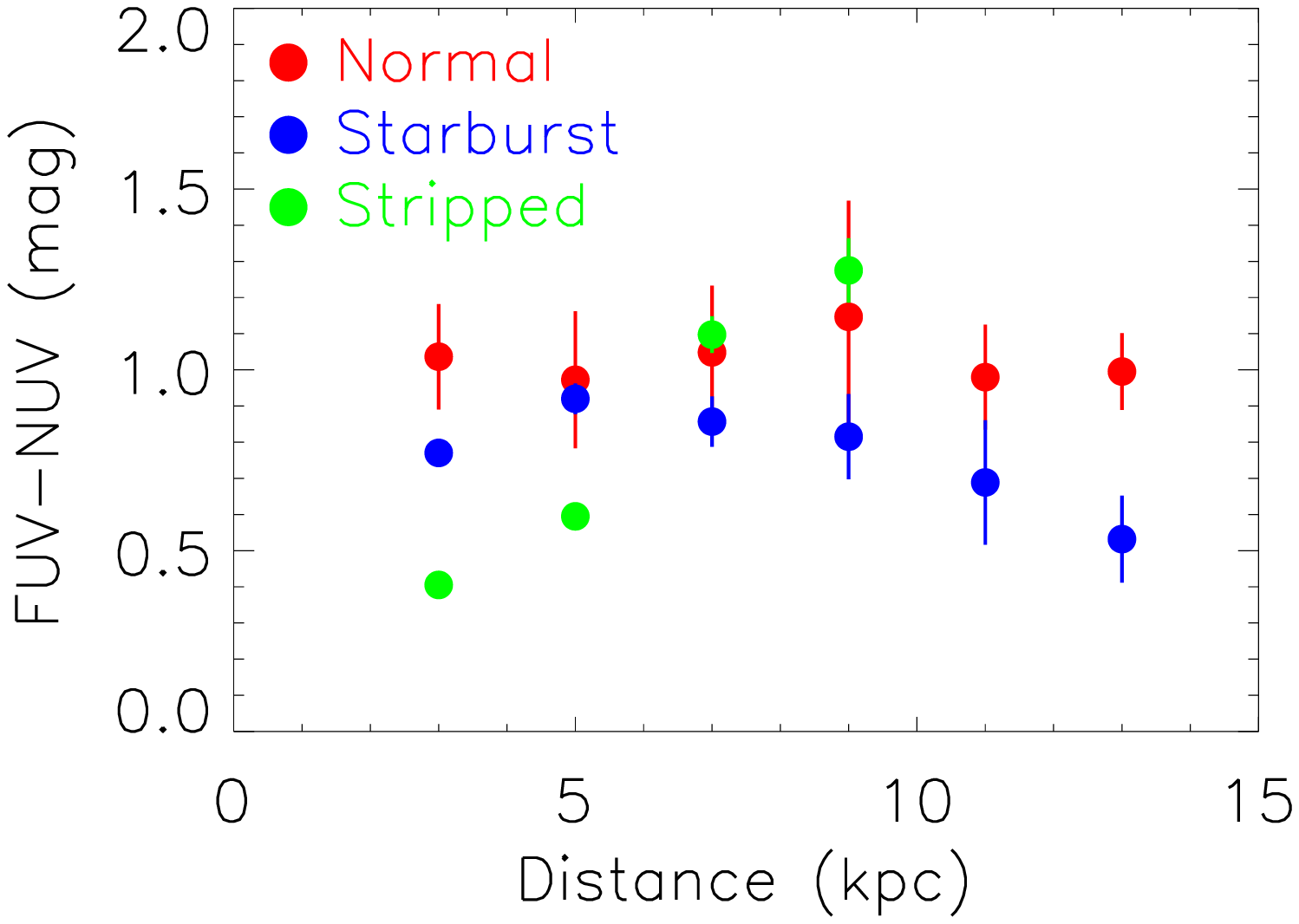}
\caption{\scriptsize FUV$-$NUV color as a function of height for
  composites of the normal, starburst, and stripped galaxies in our sample.}
\label{figure.fuvnuv_height}
\end{center}
\end{figure}

To improve the signal, we measured the FUV$-$NUV color as a function
of height in composites of the normal, starburst, and stripped
galaxies (of which NGC~4522 and NGC~4330 have \Galex{} data). The
composite measurements are shown in
Figure~\ref{figure.fuvnuv_height}. The
data were binned in boxes 2\,kpc wide, with the first box having a
central height of 3\,kpc from the midplane. The projected distances
from the midplane for each measurement were used to assign each
measurement to the appropriate bin, and the galaxies were projected to
a common distance of 20\,Mpc for the purpose of computing the
magnitudes. 
Figure~\ref{figure.fuvnuv_height} reinforces the conclusions from the
individual galaxies: the starburst halos become bluer with height, the
normal galaxies have an approximately constant color, and the stripped
galaxies are strikingly blue near the disk and similar to the normal
galaxies at larger heights. 
However, some variation in the color with
height is seen in a few individual normal galaxies. More data are necessary
to determine if this is similar to what is seen in the starbursts.

\subsection{Correlation Analysis for Average Properties}

\begin{turnpage}
\begin{deluxetable*}{lcccccccccccccccc}
\tablenum{6}
\tabletypesize{\scriptsize}
\tablecaption{Diffuse Halos}
\tablewidth{0pt}
\tablehead{
\colhead{Name} & \colhead{FUV$-$NUV} & \colhead{Scale Height} & \colhead{$\chi^2_{\nu}$} & \multicolumn{4}{c}{Extraplanar Diffuse?} & \colhead{$F_{\text{FIR}}$} & \colhead{$A_{0.16}$} & \colhead{$A_{0.2}$} & 
    \multicolumn{2}{c}{$L_{\text{halo}}$} & \multicolumn{2}{c}{$L_{\text{gal}}$} & \multicolumn{2}{c}{$L_{\text{halo}}/L_{\text{gal}}$} \\
\colhead{}     & \colhead{(mag)} & \colhead{(kpc)} &  & \colhead{(H$\alpha$)} & \colhead{(X-ray)} & \colhead{(160$\mu$m)} & \colhead{(GHz)} & \colhead{} & \colhead{(mag)} & \colhead{(mag)} & 
   \colhead{(FUV)} & \colhead{(\uvmtwo{})} & \colhead{(FUV)} & \colhead{(\uvmtwo{})} & \colhead{(FUV)} & \colhead{(\uvmtwo{})} \\
\colhead{(1)} & \colhead{(2)} & \colhead{(3)} & \colhead{(4)} & \colhead{(5)} & \colhead{(6)} & \colhead{(7)} & \colhead{(8)} & \colhead{(9)} & \colhead{(10)} & \colhead{(11)} & 
    \colhead{(12)} & \colhead{(13)} & \colhead{(14)} & \colhead{(15)} & \colhead{(16)} & \colhead{(17)} 
} 
\startdata
\cutinhead{Starbursts}									
NGC0253		& 1.26$\pm$0.02		& 2.2$\pm$0.1	& 38	& Y	& Y	& Y	& Y	& 67.6	& 3.6	& 		& 6.9		& 		& 1448.2	& 			& 0.005	& 		\\
M82			& 0.87$\pm$0.02 	& 2.7$\pm$0.1	& 32	& Y	& Y	& Y	& Y	& 86.8	& 7.5	& 6.2	& 13.0		& 24.6	& 2621.6	& 3151.3	& 0.005	& 0.008	\\
NGC4631		& 0.73$\pm$0.03		& 1.8$\pm$0.1	& 1.3	& Y	& Y	& Y	& Y	& 6.4	& 1.5	& 1.2	& 11.9		& 21.8	& 612.7		& 648.2		& 0.019	& 0.035	\\
NGC3628		& 1.55$\pm$0.08		& 3.9$\pm$0.2	& 1.8	& Y	& Y	& N	& N	& 5.7	& 4.1	& 3.0	& 6.5		& 17.2	& 1347.4	& 1128.5	& 0.005	& 0.016	\\
NGC4666		& 0.90$\pm$0.08		& 7.6$\pm$0.4	& 0.3	& Y	& Y	& Y	& Y	& 3.9	& 3.7	& 3.0	& 8.2		& 12.9	& 2333.3	& 1932.9	& 0.003	& 0.007	\\
NGC3079		& 0.35$\pm$0.01		& 3.4$\pm$0.1	& 1.6	& Y	& Y	& Y	& Y	& 4.9	& 3.3	& 2.6	& 18.7		& 28.0	& 3772.2	& 3214.7	& 0.005	& 0.009	\\
NGC5775		& 0.59$\pm$0.04		& 4.5$\pm$0.3	& 2.6	& Y	& Y	& 	& Y	& 2.3	& 4.0	& 3.2	& 11.4		& 19.8	& 1887.0	& 1563.7	& 0.006	& 0.013	\\
NGC4388		& 1.02$\pm$0.06		& 2.7$\pm$0.1	& 3.5	& Y	& Y	& Y	& Y	& 1.1	& 3.0	& 2.1	& 6.0		& 14.6	& 945.9		& 878.7		& 0.006	& 0.017	\\
\cutinhead{Normal Spirals}								 				
NGC0055		& 2.8$\pm$0.6		& 1.1$\pm$0.1	& 4.2	& Y	& N	& Y	& 	& 4.0	& 0.5	& 		& 0.5		& 		& 65.7		& 			& 0.007	& 		\\
NGC0891		& 0.33$\pm$0.09		& 1.5$\pm$0.2	& 0.8	& Y	& Y	& Y	& Y	& 6.8	& 5.0	& 4.3	& 6.9		& 11.4	& 1088.3	& 1043.1	& 0.006	& 0.010	\\
NGC2683		& 2.0$\pm$0.4		& 2.7$\pm$0.2	& 0.9	& N	& N	&	& N	& 1.1	& 2.3	& 1.7	& 3.6		& 5.8	& 203.6		& 219.7		& 0.018	& 0.025	\\
NGC4517		& 1.5$\pm$0.2		& 3.4$\pm$0.2	& 1.9	& 	& 	& Y	& 	& 0.7	& 1.9	& 		& 3.1		& 		& 152.2		& 			& 0.020	& 		\\
NGC4565		& 					& 2.0$\pm$0.3	& 0.7	& N	& N	& N	& N	& 1.2	& 		& 		& 			& 		& 			& 			&		& 		\\
NGC4096		& 1.8$\pm$0.3		& 4.3$\pm$0.3	& 0.7	& N	& 	& Y	& N	& 0.9	& 2.0	& 1.5	& 1.9		& 7.9	& 277.6		& 312.3		& 0.007	& 0.024	\\
NGC4313		& 1.0$\pm$0.3		& 6.6$\pm$3.3	& 2.7	& 	& 	&	& 	& 0.2	& 3.9	& 		& 1.3		& 		& 73.7		& 			& 0.021	& 		\\
NGC3623		& 3.$\pm$3.			& 4.$\pm$1.		& 3		& N	& N	&	& N	& 0.6	& 2.3	& 1.4	& 0.4		& 3.3	& 169.9		& 209.4		& 0.002	& 0.015	\\
NGC5907		& 0.74$\pm$0.09		& 6.4$\pm$0.2	& 1.3	& N	& N	& Y	& N	& 1.7	& 2.5	& 2.0	& 7.1		& 13.8	& 816.6		& 829.1		& 0.009	& 0.016	\\
NGC4216		& 1.6$\pm$0.3		& 3.4$\pm$0.3	& 0.9	& N	& 	& N	& N	& 0.3	& 2.6	& 		& 2.7		& 		& 160.1		& 			& 0.017	& 		\\
NGC4607		& 1.2$\pm$0.4		& 3.7$\pm$1.3	& 1.0	& 	& 	& N	& 	& 0.4	& 4.7	& 		& 0.8		& 		& 202.1		& 			& 0.004	& 		\\
NGC4522		& 0.55$\pm$0.04		& 1.9$\pm$0.1	& 1.8	& Y	& 	& Y	& 	& 0.2	& 1.6	& 		& 6.0		& 		& 139.4		& 			& 0.043	& 		\\
NGC0134		& 					& 1.9$\pm$0.3	& 0.7	& 	& 	& 	& 	& 2.8	& 		& 2.5	&			& 10.1	& 			& 1548.1	&  		& 		\\
NGC4157		& 0.8$\pm$0.1		& 4.7$\pm$0.5	& 2.1	& N	& 	& 	& Y	& 2.0	& 3.6	& 		& 6.3		& 		& 1171.6	& 			& 0.005	& 		\\
ESO358		& 1.2$\pm$0.1		& 5.3$\pm$0.9	& 0.1	& 	& 	&	& 	& 0.5	& 2.8	& 		& 2.6		& 		& 296.9		& 			& 0.009	& 		\\
NGC4217		& 1.2$\pm$0.2		& 3.1$\pm$0.4	& 0.1	& Y	& Y	& Y	& Y	& 		& 		&		& 7.6		& 		& 			& 			&  		& 		\\
NGC4330		& 0.51$\pm$0.03		& 1.1$\pm$0.1	& 1.6	& 	& 	&	& 	& 0.1	& 1.7	& 1.4	& 5.1		& 6.9	& 101.7		& 116.5		& 0.050	& 0.055	\\
NGC3044		& 0.2$\pm$0.1		& 3.8$\pm$0.7	& 0.8	& Y	& Y	&	& Y	& 1.1	& 2.4	& 1.9	& 6.7		& 6.0	& 998.1		& 1011.2	& 0.007	& 0.006 \\
NGC5170		& 3.$\pm$2.			& 7.9$\pm$0.5	& 0.8	& N	& N	&	& 	& 0.2	& 1.5	& 		& 3.6		& 		& 269.1		& 			& 0.013	& 		
\enddata
\tablecomments{\label{table.derived} Cols. (1) Name (2) FUV$-$NUV color averaged over the halo (3-4) Best-fit scale height to combined data from all wavebands and reduced $\chi^2$
(5-8) Is there extraplanar diffuse emission? (9) FIR flux used to determine galactic extinction in units of $10^{-9}$\,erg\,s$^{-1}$\,cm$^{-2}$ (10-11) Estimated extinction in the FUV and \uvmtwo{} bands from \citet{buat99}
(12-13) Halo luminosity density in units of $10^{25}$~erg\,s$^{-1}$\,Hz$^{-1}$ (13-14) De-reddened galaxy luminosity density in units of $10^{25}$~erg\,s$^{-1}$\,Hz$^{-1}$ 
(15-16) Halo-to-galaxy luminosity ratio using de-reddened galaxy luminosity.}
\tablerefs{The presence of extraplanar diffuse emission was
  determined from the literature. H$\alpha$: \citet{rossa03a}, X-ray:
  \citet{strickland04,tullmann06,li13}, Radio: \citet{wiegert15} 
}
\end{deluxetable*}
\end{turnpage}

We compared the total halo fluxes, FUV$-$NUV colors, and the scale
heights to the following galaxy parameters: the UV luminosity of the
galaxy ($L_{\text{gal}}$, corrected for internal extinction), the
star-formation rate (SFR) and specific SFR (sSFR), the
H$\alpha$ luminosity of the galaxy ($L_{\text{H}\alpha}$), the stellar
mass ($M_{*}$), the rotation velocity ($v_{\text{rot}}$), the
morphological type code \citep[$T$;][]{devaucouleurs91}, the
inclination ($i$), and the distance from Earth ($d$). 

$L_{\text{H}\alpha}$, $v_{\text{rot}}$, $T$, $i$, and $d$ come from
NED and HyperLeda values. Some of the H$\alpha$ fluxes include some
flux from the neighboring [N{\sc ii}] line. We compute $M_*$ from the
$K$-band luminosity from the 2MASS Extended Source Catalog and the
mass-to-light ratio from \citet{bell01}, for which we use $B-V$ colors
from HyperLeda. The SFR is based on the relation from
\citet{kennicutt98}, which uses far-infrared (FIR) fluxes measured
with the IRAS satellite. Fluxes exist for all but one galaxy
(NGC~4217). The values we use are reported in
Table~\ref{table.sample}.

We estimated $L_{\text{gal}}$ in each filter by ``de-reddening'' the
measured flux following \citet{buat99}, who found a relation for the
UV extinction based on the far-infrared (FIR) and measured UV fluxes.
The relation is best calibrated for the FUV and \uvmtwo{} filters, so
we focus on these filters. We further correct $L_{\text{gal}}$ for the
light that would escape the galaxy assuming an extinction out of the
disk of 0.5\,mag in the $B$ band \citep{calzetti01}.  
We convert this
to the extinction in the FUV and \uvmtwo{} filters using a
modified \citet{calzetti00} law from \citet{kriek13}, who derived
extinction curves for composite galaxy SEDs. Following their
scheme, we adopt an attenuation slope of $\delta=-0.2$ for the starbursts
and $\delta=0.0$ for the normal galaxies, and a 2175\AA\ bump strength of
$E_b = 0.5$ or $1.0$ for starbursts and normal galaxies, respectively.
These are both more similar to the \citet{calzetti00} law than the traditional
Milky Way extinction law towards \citep{weingartner01}. This leads to
$A_{\text{FUV}}/A_B = 2.05$ and $A_{\text{uvm2}}/A_B = 1.96$ for the starburst
galaxies, whereas $A_{\text{FUV}}/A_B = 2.5$ and $A_{\text{uvm2}}/A_B = 2.2$
for the normal galaxies. These parameters are appropriate for star-forming
galaxies, but the analysis below is insensitive to the exact $\delta$ or
$E_b$, or whether $L_{\text{gal}}$ includes attenuation along the minor axis.
The measured $L_{\text{gal}}$, the estimated extinction, and the corrected
values are given in Table~\ref{table.derived}.

We used Spearman's rank correlation with a significance threshold of
$p=0.05$ to search for correlations between galaxy and halo
properties. A summary of the findings is given in
Table~\ref{table.correlations}. Here we describe the results and give the 
correlation coefficient ($R$) and $p-$value for correlations with $p < 0.05$.
As we tested for correlations between a range of measurements, 
we adopted the \citet{benjamini95} procedure to control for false positives, 
also using a threshold of $\alpha=0.05$. There were 61 total comparisons.
Correlations with $p$-values that remain significant for the revised threshold
are noted in Table~\ref{table.correlations} and discussed below.

The halo luminosity $L_{\text{halo}}$ is strongly correlated with 
$L_{\text{gal}}$, with $R=0.87$ and $p < 10^{-8}$ in FUV
and $R=0.79$ and $p = 8.3\times 10^{-4}$ in \uvmtwo{}. It is also
correlated with the SFR in both ($R=0.82$ and $p=8.5\times 10^{-7}$ for
the FUV and $R=0.77$ with $p=0.0012$ in \uvmtwo{}), and with the specific
SFR ($R=0.64$ and $p=0.0016$ in FUV and $R=0.59$ and $p=0.033$ in \uvmtwo{}).
$L_{\text{H}\alpha}$ 
is closely related to the SFR, but whereas there is a significant
correlation between $L_{\text{halo}}$ and $L_{\text{H}\alpha}$ in the FUV
($R=0.62$, $p=0.0041$) there is no significant correlation for \uvmtwo{} ($p=0.12$).
$L_{\text{halo}}$ is correlated with morphological type for \uvmtwo{}
($R=0.58$, $p=0.030$) but not for the FUV ($p=0.12$). Overall, the $p$-values are smaller
for the FUV, possibly because there are 24 FUV measurements and 15 \uvmtwo{}
measurements. There is no other significant correlation between $L_{\text{halo}}$
and the other parameters in either filter. 

$L_{\text{halo}}$ is also correlated with the FUV$-$NUV color for the FUV
band ($R=0.52$, $p=0.00076$), but not \uvmtwo{} ($p=0.11$). The FUV-NUV color is
significantly correlated with the specific SFR ($R=0.48$, $p=0.028$) but
not the SFR ($p=0.076$), nor with $L_{\text{gal}}$ in either FUV or \uvmtwo{}
($p=0.051$ and $p=0.38$, respectively). There are no other significant
correlations between FUV$-$NUV and other galaxy properties.

In addition to the FUV $L_{\text{halo}}$, the FUV$-$NUV color is 
correlated with the FUV $L_{\text{gal}}$ ($R=0.41$, $p=0.048$). It is
not correlated with other halo luminosities. FUV$-$NUV is also
marginally correlated with sSFR ($R=0.48$, $p=0.03$), but the
correlation with SFR is not significant. There are no other
significant correlations between FUV$-$NUV and the galaxy properties.

We also compared $L_{\text{halo}}/L_{\text{gal}}$ to the galaxy
properties for the FUV and \uvmtwo{}
filters. $L_{\text{halo}}/L_{\text{gal}}$ is correlated
with the SFR ($R=0.42$ and $p=0.042$ for the FUV, and $R=0.59$ with
$p=0.027$ for \uvmtwo{}). There are no other significant correlations
with other properties. 

The scale height $h$ is not correlated with any galaxy or halo property
except $d$, where $R=0.47$ and $p=0.012$. This is almost entirely due
to NGC~5170 ($h=7.9$\,kpc), which is near the distance cutoff and has
an unusually large value.

Applying the Benjamani-Hochberg procedure for $\alpha=0.05$ and 61 tests,
we find that the following correlations are significant: in both 
FUV and \uvmtwo{} $L_{\text{halo}}$ is correlated with $L_{\text{gal}}$
and with the SFR, and $L_{\text{halo}}$ is correlated with FUV$-$NUV, specific
SFR, and $L_{\text{H}\alpha}$ in the FUV only. The other correlations
with $p<0.05$ described above are not significant at this threshold.

\begin{deluxetable*}{lcccccc}
\tablenum{7}
\tabletypesize{\scriptsize}
\tablecaption{UV Halo Correlations}
\tablewidth{0pt}
\tablehead{
\colhead{Quantity} & \multicolumn{2}{c}{$L_{\text{halo}}$} & \multicolumn{2}{c}{$L_{\text{halo}}/L_{\text{gal}}$} & \colhead{$h$} & \colhead{FUV$-$NUV} \\
\colhead{} & \colhead{(FUV)} & \colhead{(\uvmtwo{})} & \colhead{(FUV)} & \colhead{(\uvmtwo{})} & \colhead{(kpc)} & \colhead{(mag)} \\
\colhead{(1)} & \colhead{(2)} & \colhead{(3)} & \colhead{(4)} & \colhead{(5)} & \colhead{(6)} & \colhead{(7)}
}
\startdata
$L_{\text{gal,FUV}}$     & 0.87\tablenotemark{a} & -    & N    & -    & N    & 0.41\tablenotemark{a} \\
$L_{\text{gal,uvm2}}$    & -    & 0.79\tablenotemark{a} & -    & N    & N    & N    \\
$h$                      & N    & N    & N    & N    & -    & N    \\
$L_{\text{gal,H}\alpha}$ & 0.62\tablenotemark{a} & N    & N    & N    & N    & N    \\
SFR(IR)                  & 0.82\tablenotemark{a} & 0.77\tablenotemark{a} & 0.42 & 0.59 & N    & N    \\
$M_*$                    & N    & N    & N    & N    & N    & N    \\
SFR(IR)/$M_*$            & 0.64\tablenotemark{a} & 0.59 & N    & N    & N    & 0.48 \\
$v_{\text{rot}}$         & N    & N    & N    & N    & N    & N    \\
$T$                      & N    & 0.58 & N    & N    & N    & N    \\
$i$                      & N    & N    & N    & N    & N    & N    \\
$d$                      & N    & N    & N    & N    & 0.47 & N    
\enddata
\tablecomments{\label{table.correlations} Correlations between halo quantities and galaxy and halo properties.
For significant correlations ($p<0.05$) the Spearman ranked correlation coefficient is given. Non-significant
correlations are marked by an `N'. A dash (`-') indicates no measurement.  
Cols. (1) Quantity (2-3) Halo luminosity (4-5) Ratio of halo luminosity to de-reddened galaxy luminosity
(6) Scale height of the UV halo (7) Average FUV$-$NUV color in the halo. See text for more a more detailed
description.}
\tablenotetext{a}{These correlations survive the Benjamini-Hochberg test with $\alpha=0.05$ for 61 tests.}
\end{deluxetable*}

In summary, $L_{\text{halo}}$ is related to $L_{\text{gal}}$ and some the
metrics of star formation, as well as the halo color. However, the scale
height has no (strong) dependence on the halo or intrinsic galaxy properties.

\subsection{Average Properties and Morphology}

For most galaxies in the \Galex{} and \Swift{} archive it is not
possible to examine the morphology in detail because of low $S/N$ or
large distance. Thus, we compare the average properties
$L_{\text{halo}}$, FUV$-$NUV, and $h$ to the UV maps in
Figures~\ref{figure.gallery_normal} and
\ref{figure.gallery_starburst}. Overall, we find that averaging
obscures the presence of halo components such as winds,
filaments, or bright emission near the disk, so that these properties
are not useful metrics of the halo morphology. 

Apart from the clear divide between normal and starburst galaxies,
$L_{\text{halo}}$ appears unrelated to the halo morphology, except in
stripped galaxies. For
example, NGC~5907 has a very patchy, low $S/N$ halo in
Figure~\ref{figure.gallery_normal}, but its $L_{\text{halo}}$ is among
the highest of the normal galaxies. Since we know that
$L_{\text{halo}}$ is uncorrelated with $h$ it does not provide information about the halo
structure. Likewise, the most spectacular UV halos around starburst
galaxies (M82, NGC~253, NGC~4631, and NGC~3079) have similar
$L_{\text{halo}}$. The stripped galaxies (NGC~134, NGC~4522, and
NGC~4330) have unusually high $L_{\text{halo}}/L_{\text{gal}}$
(5-10\% instead of the typical 0.5-2\%). As discussed earlier, this
is largely attributable to extraplanar star formation. 

We also find no connection between the halo morphology and the average
FUV$-$NUV color, except insofar as bluer halos are
more structured in our sample because they predominantly belong to
starburst galaxies. 
Likewise, the scale height appears to be
independent of radial concentration, although in starburst or stripped
galaxies the scale height can change with galactocentric radius.
The stripped galaxies cannot be
distinguished from the rest of the sample by their average FUV$-$NUV
color and are only barely distinguishable by scale height; it is not
likely that these metrics could identify stripped galaxies in a wider
sample. 

\subsection{Differences with HKB14}

In \citetalias{hk14} we measured fluxes without the PSF-wing
correction and used these fluxes to find scale heights for several
galaxies and measure correlations between some of the same quantities
as above. The PSF-wing contamination is strongest near the disk, so we
might expect $h$ to increase when the spurious flux is removed, but
this is not generally the case. However, the correlations that we find
between $L_{\text{halo}}$ and $L_{\text{gal}}$, the H$\alpha$
luminosity, and the FIR SFR are stronger after $L_{\text{halo}}$ is
corrected for the PSF-wing contamination. For example, in the FUV the
correlation between $L_{\text{halo}}+L_{\text{PSF contam}}$ and
$L_{\text{gal}}$ is $R=0.41$ whereas between $L_{\text{halo}}$ and
$L_{\text{gal}}$ it is $R=0.87$. The
lower fluxes also change the SEDs, but the only trend reported in
\citetalias{hk14} that is truly artificial is the uniform rise in the
FUV$-$NUV color with height; we find in most galaxies no change with
height. 

\citet{shinn15} investigated the role of PSF-wing contamination in
several galaxies included in \citetalias{hk14}, and found that in two
cases (NGC~24 and IC~5249) the UV halo that we reported appears to be
entirely artificial. They used the \Galex{} images exclusively, and
extrapolated the PSF wings from the PSF provided by the \Galex{}
calibration team. Using the PSFs described in
Section~\ref{section.psf}, we agree with their findings for the
\Galex{} images. For the considerably deeper \Swift{} images, we find
a faint UV halo around NGC~24 (which is not edge-on) and no UV halo
above 2\,kpc around IC~5249. IC~5249 is seen edge-on has a very thin
optical disk, so it is possible that any astrophysical UV halo exists
primarily below this. Based on these results and the detections in
our present sample, UV halos are ubiquitous but may not be universal
(or universally detectable with current instruments). 

\section{Discussion}
\label{section.discussion}

\subsection{The Diffuse UV Light is Probably a Reflection Nebula}

Prior studies have argued that the UV halos are eRN on the basis that the
light is too bright to come from shock-heated or photoionized gas
\citep{hoopes05}, that line emission from starburst halos is polarized in a way that is consistent
with dust scattering \citep{yoshida11}, that it can be successfully
modeled by Monte Carlo radiative transfer scattering models
\citep{seon14,shinn15}, that it is too blue to originate in the
stellar halo \citepalias{hk14}, and that they are coincident with
eDIG, winds, and other outflow tracers.

Our results further support this scenario:
\begin{enumerate}
\item The UV halos are truly diffuse, as determined by comparison with
  existing HST data, and they trace filamentary morphology seen at
  other wavelengths.
\item UV halos are a broadband phenomenon, with similar morphology
  seen from 1500-2600\AA.
\item $L_{\text{halo}}$ is strongly correlated with $L_{\text{gal}}$
  after de-reddening the galaxy flux, and it is strongly correlated
  with the SFR.
\item Extraplanar UV light is most visible above regions of active
  star formation as seen in H$\alpha$, X-ray, and radio continuum
  data, but the presence or prominence of the UV halo does not depend
  on the presence of extraplanar H$\alpha$, X-rays, or radio
  continuum. This rules out an emission nebula as the source.
\item Starburst galaxies have more luminous UV halos, but not higher
  $L_{\text{halo}}/L_{\text{gal}}$ values or scale heights.
\item Starburst winds are visible through their smaller FUV$-$NUV
  colors, which appear to be superimposed on a thick disk similar to
  (although brighter than) that around normal galaxies.
\item The UV halo flux, color, and scale height are not significantly
  correlated with galaxy properties that are not closely connected to
  star formation.
\end{enumerate}
The strong connection between $L_{\text{halo}}$ and $L_{\text{gal}}$ and
the weak connection between $L_{\text{halo}}/L_{\text{gal}}$ and the
presence of outflows strongly suggests that UV halos are eRN. This also
explains their broadband visibility and ubiquity, as non-ionizing UV
photons can escape even if the disk is not porous. 
If UV halos are eRN, then they indicate that dust is widespread in the
halos of spiral galaxies of all types. This implies that dust is
long-lived in galaxy halos to the extent that it exists around galaxies
without strong outflows and has a similar scale height as the dust around
starburst galaxies. In the remainder of this section, we assume that the
UV halos are eRN and examine some basic properties (however, we defer a
detailed analysis of the UV halo SEDs to Paper~III). We then discuss the
results in the context of other work on extraplanar dust.

\subsection{Dust Mass}

We estimate the dust mass using Monte Carlo radiative transfer (MCRT)
models based on the model described in \citet{wood01} and
\citet{whitney11}, which incorporates the Henyey-Greenstein functions
for the angular dependence of dust scattering and guarantees that each
photon scatters once. These models are meant to be first-order
estimates and are not as carefully constructed as other MCRT models of
eRN such as in \citet{shinn15} or \citet{baes16}, but they are
nonetheless useful.

\begin{deluxetable*}{lccccccccc}
\tablenum{8}
\tabletypesize{\scriptsize}
\tablecaption{MCRT Models \& 160\,$\mu$m Masses}
\tablewidth{0pt}
\tablehead{
\colhead{} & \multicolumn{2}{c}{Thick Disk} & \multicolumn{3}{c}{Milky Way Dust} & \multicolumn{3}{c}{SMC Dust} &  \\
\colhead{Name} & \colhead{$R_0$} & \colhead{$z_0$} & \colhead{$G/D$} & \colhead{$\chi^2_{\nu}$} & \colhead{$M_{\text{ext}}$} & \colhead{$G/D$} & \colhead{$\chi^2_{\nu}$} & \colhead{$M_{\text{ext}}$} & \colhead{$M_{160\mu\text{m}}$}  \\
\colhead{} & \colhead{(kpc)} & \colhead{(kpc)} & \colhead{} & \colhead{} & \colhead{($10^6 M_{\odot}$)} & \colhead{} & \colhead{} & \colhead{($10^6 M_{\odot}$)} & \colhead{($10^6 M_{\odot}$)} \\
\colhead{(1)} & \colhead{(2)} & \colhead{(3)} & \colhead{(4)} & \colhead{(5)} & \colhead{(6)} & \colhead{(7)} & \colhead{(8)} & \colhead{(9)} & \colhead{(10)} 
}
\startdata
NGC 891		& 12 & 1.9 & $60^{+80}_{-20}$ 		& 1.0	& $6\pm3$ 				& $50^{+120}_{-10}$		& 0.9	& $7^{+2}_{-5}$		  & 3.2 \\
NGC 4631	& 8  & 2.3 & $450^{+150}_{-50}$		& 4.6	& $1.2^{+0.3}_{-0.1}$ 	& $700^{+400}_{-200}$	& 4.1	& $0.7\pm0.3$ 		  & 0.7 \\
NGC 5775	& 12 & 3.0 & $250^{+50}_{-50}$		& 9.9	& $4.0^{+0.6}_{-0.8}$ 	& $400^{+100}_{-50}$	& 9.3	& $2.3^{+0.4}_{-0.5}$ & -   \\
NGC 5907	& 17 & 3.0 & $400^{+250}_{-150}$	& 1.1	& $2.5^{+1.5}_{-0.8}$	& $300^{+200}_{-150}$	& 1.4	& $3.3^{+3.1}_{-1.3}$ & 6.1 
\enddata
\tablecomments{\label{table.mcrt} 
Cols. (1) Name (2-3) Scale lengths for the thick exponential disk from matching 
\ion{H}{1} profiles (4-6) Best-fit $D/G$, reduced $\chi^2$, and implied extraplanar 
dust mass above 2\,kpc for models with Milky Way dust. The error bars are statistical
from the MCRT fitting only and underestimate the true uncertainty. (7-9) Best-fit $D/G$, 
reduced $\chi^2$, and extraplanar dust mass for SMC dust. The fits were performed
using a dust opacity and scattering albedo at 2000\AA\ and \uvwtwo{} data. See
text for details. (10) Lower bound to dust mass above 2\,kpc from 160\,$\mu$m fluxes.
See text for details.
}
\end{deluxetable*}

In principle, the dust mass can be measured directly from the 
$L_{\text{halo}}/L_{\text{gal}}$ ratio, since
\begin{equation}
L_{\text{halo},\nu} = L_{\text{gal},\nu} (1-e^{-\tau_{\nu} \varpi_{\nu}})
\end{equation}
where $\varpi_{\nu}$ is the scattering albedo and $\tau_{\nu} = 1.086
N_{\text{dust}} \sigma^{\text{ext}}_{\nu}$ is the optical
depth. $\sigma^{\text{ext}}_{\nu}$ is the extinction cross-section and
$N_{\text{dust}}$ the column. The extinction through the halo is low
($\tau_{\nu} \ll 1$), so a single-scattering approximation is
reasonable. As described in \citetalias{hk14},
$\sigma^{\text{ext}}_{\nu}$ and $\varpi_{\nu}$ can be determined from
the shape of the SED, so the normalization is related to a
characteristic column density from the disk through the halo. However,
the measured flux at a given height above the disk cannot be
straightforwardly interpreted as a dust column because the light
source is not behind the dust. Thus, one must also adopt a geometric
model for the halo dust and the disk emission, for which we use MCRT. 

The MCRT model needs input distributions for the emission and the
scattering/absorbing medium, as well as the viewing angle and the
scattering cross-section ($\sigma^{\text{ext}}_{\nu} \times
\varpi_{\nu}$). 

\begin{figure}
\begin{center}
\includegraphics[width=0.5\textwidth]{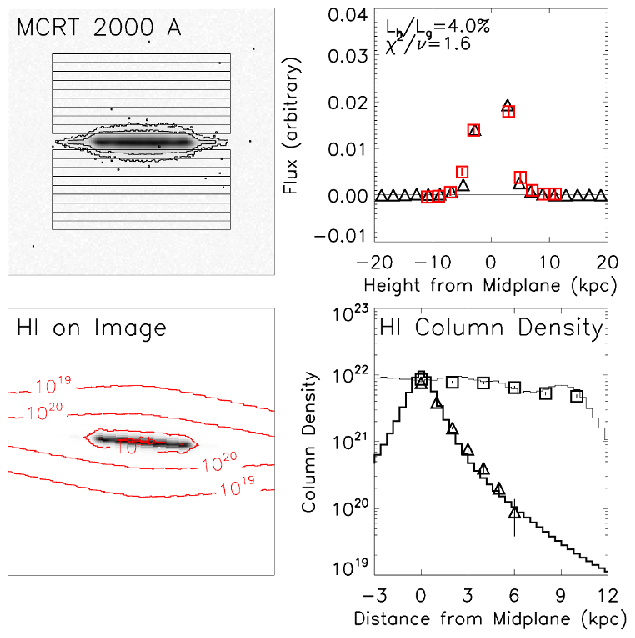}
\caption{\scriptsize The best-fit MCRT model to NGC~4631 for Milky Way-like
dust. \textit{Top Left}: MCRT image at 2000\AA\ with 3 and 6$\sigma$ contours
overlaid, along with flux measurement boxes. The average noise level
from the \uvwtwo{} image was added to create the contours and show the
``observable'' extent of the halo. \textit{Top Right}: Measured 
\uvwtwo{} fluxes from NGC~4631 compared to values measured from the MCRT model
in the boxes at left. Fluxes in both cases are normalized to the measured galaxy
flux. \textit{Bottom Left}: Model \ion{H}{1} contours overlaid on the MCRT
image, which is rotated to the position angle of NGC~4631. \textit{Bottom Right}:
Measured \ion{H}{1} column densities overlaid on our \ion{H}{1} model. The thick
black line shows the profile along the minor axis, while the thin line shows 
the major axis.
}
\label{figure.mcrt}
\end{center}
\end{figure}

We assume that the dust is embedded in the neutral medium and model
the gas density using several components: a thin exponential disk,
simple logarithmic spiral arms, and a thick exponential disk that
represents the halo. Each exponential disk takes the form $\rho(R,z) =
\rho_0 e^{-R/R_0}e^{-|z|/z_0}$, where $R$ is the galactocentric radius
in the plane and $z$ is vertical height above the plane. We constrain $\rho_0$, $R_0$ and
$z_0$ by using high resolution \ion{H}{1} maps, so we only model
galaxies where we have such maps in hand: NGC~891, NGC~4631, NGC~5775,
and NGC~5907. $R_0$ in the halo is unconstrained, so we tie it to the
value of $R_0$ in the disk.
The spiral arms are described by a polar equation
$r = a e^{b\theta}$, where $a$ is the normalization and $b$ the rate
of growth (the parametric equations for a
Cartesian grid are $x = r\cos\theta$ and $y=r \sin \theta$). We choose
$a=7$ and $b0.28$ and compute $x$ and $y$ for $\theta \in [0,3\pi]$ with
a minimum radius $R=3$\,kpc. To add thickness, these spiral arms are
convolved with a 3D Gaussian kernel with $\sigma=8$~pixels and an amplitude
of $0.1 \rho_0$, which is added to the underlying exponential disk. We
fixed the parameters based on the extent of typical spiral arms, but for
an edge-on viewing angle our results are insensitive to a wide range of $a$,
$b$, or spiral arm width. 
The lower panels of
Figure~\ref{figure.mcrt} show the model for NGC~4631; we do not
attempt to reproduce large-scale filaments or warps.

The light source consists of a very thin exponential disk with the
same scale length as the thin gas disk and light from the midplane in the
spiral arms,
since we assume that UV light comes from young stars near the midplane
and in the arms. Our results are not sensitive to the shape or the
location of the spiral arms because we measure the halo fluxes by
summing over $R$ (described below). 

Paper~III will be devoted to modeling the SED and
determining the dust size and composition, which will yield 
$\sigma^{\text{ext}}_{\nu}$ and $\varpi_{\nu}$. Here we adopt a
simpler approach by using the values from two dust models from
\citet{weingartner01}: dust in the bar of the Small Magellanic Cloud
(SMC), and dust in the Milky Way for sightlines where the extinction
law $R_V = 3.1$. The choice of these models is motivated by the
observation that the halo dust at larger radii appears to have an
SMC-like extinction curve \citep{menard10}, and that many galaxy disks
have extinction curves similar to the Milky Way. As we shall see, the
inferred dust masses for each model are similar.

The density and emission model is gridded on a $201\times 201\times
201$ cube with a pixel scale of 2.5\,pixels\,kpc$^{-1}$, and we model
the galaxy at 2000\AA\ only (where the corrected $L_{\text{gal}}$ is
most reliable). The viewing angle is matched to the inclination of
each galaxy, and we apply the dust model by assuming that the extinction
within the disk follows the \citet{kriek13} prescription with $\delta=0.0$
and $E_b = 1.0$, with a gas-to-dust ratio ($G/D$) of 100, which is typical
for spiral disks. 
Then, for a
given halo $G/D$ we simulate the light scattered into the line of
sight above 2\,kpc from the midplane and compare to the
observations. 
We repeated this procedure for a grid of $G/D \in [50,1500]$
with spacing of 10 between $G/D = 50$ to 100, and a spacing of 50 thereafter. 
The 1$\sigma$ acceptable range shown in Table~\ref{table.mcrt} is
based on $\Delta \chi^2$ on this grid.
The best-fit halo $G/D$ then implies a dust mass. For
each simulation, we use $10^7$ scattered photons (we found for several
test cases that results do not differ when using $10^8$ photons). 

For each $G/D$ we simulated an image and measured the halo light using
long boxes parallel to the midplane that are analogous to those used
to measure real UV halos. This binning smooths over differences
between spiral arm models that are not well constrained by the
data. We normalized the real and simulated fluxes to the projected
galaxy luminosity, and found the best-fit $G/D$ using the $\chi^2$
statistic. The best fit is often not a good fit, but the simulated
halos reasonably reproduce the observed scale heights and
$L_{\text{halo}}/L_{\text{gal}}$. 

As an example, we show the fit to NGC~4631 in
Figure~\ref{figure.mcrt}. The top panels show the outcome of the MCRT
simulation. In the upper left panel, noise comparable to that in the
\uvwtwo{} image was added to show how far the simulated halo could be
detected in the UVOT images (3 and 6$\sigma$ contours are shown). The
morphology of the observed halos is more complex than the symmetric,
disky morphology that we obtain by construction, but since
$L_{\text{halo}}$ and the scale height $h$ are not morphological
indicators (Section~\ref{section.fluxes}), the MCRT models are
reasonable proxies for obtaining $M_{\text{dust}}$ \citep{seon14}. The
top right panel shows the best-fit model compared to the
measurements. 
The best-fit $G/D$ values are $G/D =
450^{+150}_{-50}$ for Milky Way dust and $G/D = 700^{+400}_{-100}$ for
SMC dust, which implies a dust mass above 2\,kpc of $M_{\text{dust}} =
1.2^{+0.3}_{-0.1} \times 10^6 M_{\odot}$ for Milky Way dust and
$M_{\text{dust}} = 0.7\pm0.3 \times 10^6 M_{\odot}$ for SMC
dust. 
We emphasize that we do not know whether SMC or Milky Way dust
is closer to the true halo composition, but the similarity in
masses suggests that $M_{\text{dust}}$ is insensitive to the dust
model for a plausible range of models. The bottom panels of
Figure~\ref{figure.mcrt} show the \ion{H}{1} model compared to the
data. 

In Table~\ref{table.mcrt} we give the best-fit $G/D$, $\chi^2$, halo
component parameters, and dust mass above $|z|>2$\,kpc for the four
galaxies we considered.  We expect the best-fit $G/D$ to be similar
between filters, but with the current data this is not a useful test
because the true extinction curve is not known, and changes in $G/D$
are degenerate with changes in the incident spectrum.  We caution that
there could be systematic shifts in $G/D$ due to our choice of
geometric model, the assumption that the dust is hosted by the neutral
medium, the $G/D$ adopted in the disk, and the scattering recipe in
the MCRT code. Even if all of these are correct or unimportant, we
expect the MCRT $M_{\text{dust}}$ values to be lower limits to the halo dust
in this region because shallower profiles are allowed by the
\ion{H}{1} data and could not be constrained by our UV measurements.

Bearing these caveats in mind, it is nonetheless worth noting that the
MCRT $M_{\text{dust}}$ masses are consistent with extrapolating the
spherically symmetric \citet{menard10} radial profile (measured
between 20-1000\,kpc) to 2-10\,kpc, when we scale their average
profile based on the ratio of the mass of each of the four galaxies
considered here to their average galaxy. This suggests that there is a
smooth density transition from the disk to the halo, at least in projection.

\subsection{160\,$\mu$m Dust Mass}

The thermal emission from extraplanar dust provides a more
straightforward measurement of $M_{\text{dust}}$ because the flux is
directly related to the column density. Here we use the 160\,$\mu$m
images (as opposed to the other FIR bands) because 160\,$\mu$m is
close to the peak of the FIR SED for cool dust and because there is
more coverage of our sample with \textit{Herschel}/PACS at this
wavelength.

In most cases, the morphology of the 160\,$\mu$m emission does follow
that of the UV in the sense that it tends to be brightest where the UV
halo is also the brightest. However, the FIR emission is often not as
extended vertically or radially as the UV halo (the exceptions being
NGC~891 and NGC~4096). This could be a function of the much lower UV
background or because most of the FIR maps are shallow snapshots
obtained to measure emission from the disk, not the halo, as part of
\textit{Herschel} surveys \citep[such as the \textit{Herschel}
  Reference Survey;][]{boselli10}. 

The dust mass can be obtained from the flux, provided one knows the temperature
and grain emissivity law $\beta$. The number of grains is obtained from the
flux and these quantities:
\begin{equation}
L_{\nu} = N_{\text{dust}} 4 \pi a^2 Q_{\nu}(a) \pi B_{\nu}(T),
\end{equation}
where $Q_{\nu}(a) \approx (2\pi a \nu/c)^{\beta}$ is the dust emissivity and
$a$ the grain size. The mass implied is
\begin{equation}
M_{\text{dust}} = N_{\text{dust}} \frac{4}{3}\pi a^3 \rho_{\text{dust}},
\end{equation}
where $\rho_{\text{dust}}$ is the intrinsic density of the material. We assume
that $a=0.1$\,$\mu$m and that $\rho_{\text{dust}} = 2$\,g\,cm$^{-3}$. There is
actually a distribution of grain sizes that can be constrained with a dust
model, which we defer to Paper~III. However, we do not know the temperature
or $\beta$. $T$ is measured by fitting a modified blackbody to the far-infrared
and sub-mm SED, but almost all of the galaxies in the sample lack the data to
measure $T$ in the halo. 

For an order-of-magnitude estimate for comparison with the MCRT models, 
we assume that $\beta = 1.5$ \citep[in the literature it varies from 
1-2 in nearby galaxies; e.g.,][]{bendo03,casey12}, and use
the temperature of the cold dust measured in the disk for each
galaxy. If the dust is embedded in neutral gas, its temperature will
likely decline with height, so these $M_{\text{dust}}$ values will be
lower limits.

We measured 160\,$\mu$m fluxes above a projected height of 2\,kpc for
NGC~891, NGC~4631, and NGC~5907 from \textit{Herschel} maps (NGC~5775
does not have \textit{Herschel} data, and the angular resolution of
\textit{Spitzer}/MIPS is too poor to isolate halo flux). These fluxes
are $11.1\pm0.2$\,Jy for NGC~891, $5.4\pm0.3$\,Jy for NGC~4631, and
$3.2\pm0.4$\,Jy for NGC~5907. For temperatures we use 23\,K for
NGC~891 \citep{hughes14}, 22\,K for NGC~4631 \citep{melendez15}, and
18\,K for NGC~5907 \citep{dumke97}. Under these assumptions, the dust
masses are $M_{\text{dust}} = 3.2\times 10^6$, $0.7\times 10^6$, and
$6.1\times 10^6$ for NGC~891, NGC~4631, and NGC~5907,
respectively (Table~\ref{table.mcrt}). 

These values are very sensitive to $\beta$; if $\beta = 2$, the masses
are about ten times larger. This highlights the need for deeper FIR
data to constrain $\beta$ in the \textit{extraplanar}
dust. Considering the large uncertainties, the masses are consistent
with the MCRT values but do not provide useful constraints. 

\subsection{Dust Spatial Variation}

Some UV halos vary in FUV$-$NUV color with galactocentric radius 
(Figures~\ref{figure.fuvnuv_normal} and \ref{figure.fuvnuv_sb})
or height (Figure~\ref{figure.fuvnuv_height}). If the UV halos are
eRN, these differences arise either through changes in the dust or
the incident spectrum, although both effects may be present.
M82 provides a good example of the general behavior seen in
starbursts, namely that the eRN is bluer in the wind region
(Figure~\ref{figure.fuvnuv_sb}) and becomes bluer with height
(Figure~\ref{figure.fuvnuv_height}). 

If the dust changes in composition with height (for example, if the
fraction of silicate grains increases), the reflected spectrum could
become bluer given a constant incident spectrum. Since the optical
depth through the halo is small (as is clear from observing face-on
galaxies), this is a good approximation. On the other hand, given that
the nuclear starburst is obscured through the disk but (presumably)
not along the minor axis, the incident spectrum into the halo may
differ across the disk, leading to a projected change in the color. In
this case, sightlines vertically above the starburst region would
reflect a bluer incident spectrum than the rest of the disk. This
behavior can be generalized to \ion{H}{2} regions in the disk.

We cannot rule out either possibility, but in M82 the data suggest
that the color change in height results from a change in the dust. Let
us assume that the halo dust does not change. Then, for a partially
obscured nuclear starburst viewed edge-on, in which light can escape
along the minor axis, we expect the eRN to form bright lobes above the
starburst and near the disk. These will be visible against the rest of
the eRN. The lobes are bright near the disk because both the incident flux
and column density are higher there. Lobes of bluer color than the
rest of the eRN will also occur, with a small height offset between
the bluest and brightest region. This offset occurs because the
contribution to the eRN from the rest of the disk (with a redder
incident spectrum) is also greatest near the disk. Above the blue
lobes, the halo becomes redder as the incident flux from the starburst
declines and the reflected flux from this region competes with the
flux from the remainder of the disk. 

\begin{figure}
\begin{center}
\includegraphics[width=0.45\textwidth]{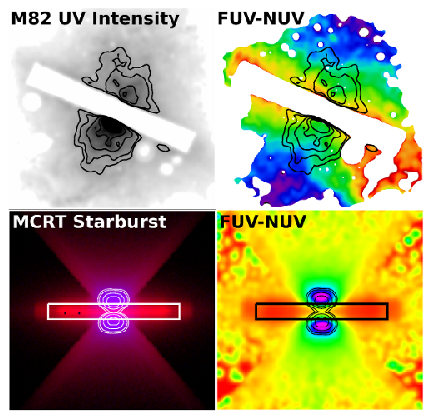}
\caption{\scriptsize \textit{Top Left}: Combined UV intensity map for M82.
\textit{Top Right}: FUV$-$NUV color map for M82. Blue or purple colors
indicate where the halo is bluer, while redder colors indicate larger
FUV$-$NUV (in magnitudes). \textit{Bottom Left}: an MCRT simulation of an
obscured starburst with a wind cone and a redder disk. The brightest
region of contours from the blue light are shown. A horizontal box indicates
the galaxy disk. \textit{Bottom Right}:
A synthetic FUV$-$NUV color map showing the halo colors with the contours from
left superimposed. We expect blue ``lobes'', which are not seen in M82.}
\label{figure.mcrt_sb}
\end{center}
\end{figure}

We confirmed these features
through MCRT models of partially obscured starbursts, as shown in
Figure~\ref{figure.mcrt_sb}. 
We tried two models: a model with the same setup as described above
except with 90\% of the emissivity clustered in the inner 1\,kpc, and
a variation of this model with a ``wind cone'' of denser material.
To examine the color we assumed that the starburst region has a
Starburst~2 spectrum and the remainder of the disk has an Sc spectrum
from the Kinney-Calzetti atlas \citep{kinney96}. We do not simulate a full
spectrum, but rather determined the proper relative luminosities for
the FUV and NUV bands. 
The cone has an opening angle of 50~degrees, a width of 0.8\,kpc, and
a density that declines exponentially with height but is uniform at each
height within the cone region. The vertical profile is $A e^{-b|z|}$,
where $A$ is 1/10 of the peak, central density that is constrained by 
the H{\sc i} data and the scale height $b = 1$\,kpc. These parameters
are motivated by the appearance of limb-brightened winds in H$\alpha$,
but are not fits to the data. We tried one additional variation in which
the interior of the cone has zero density (i.e., it lacks dust). 

In the case without the wind cone the main feature is the appearance of
the lobes described above, whereas the remainder of the eRN looks like
that for a normal disk galaxy. When the cone is present, it is brighter
than the rest of the halo and its center is bluer than average, while
the projected sides are relatively red. 
However, the presence of a cone does not change the
presence of lobes in the FUV$-$NUV color map even if the interior of
the cone is dust-free because the light is scattered by the dense cone walls.  

M82 has bright lobes near the disk and there is indeed an offset
between these lobes and the bluer region above, but there are no
corresponding lobes in the FUV$-$NUV map and the color instead
continues to become bluer with height and covers an increasingly large
fraction of the projected area (Figure~\ref{figure.mcrt_sb}). These
features cannot be explained by the simple geometric model above. The
optical depth may be higher than we assume, which would suppress lobes
in the color map, but the halo at larger heights would also be redder
as the incident spectrum would change substantially through the
halo. Some of the FUV light may come from line emission, but the same
morphology is seen in the very deep \uvwtwo{} image. The simplest
explanation of the color map is a change in the dust with height.
This could either be a physical change in a single dust component with
height or an increasing contribution from one component of dust.

In other cases, such as NGC~253 or NGC~3079, there are blue regions in
the halo near the bases of the winds that may indicate a different
light source rather than a difference in dust
(Figure~\ref{figure.fuvnuv_sb}). However, in both of these galaxies and in the 
composite starburst measurement, FUV$-$NUV color increases within a few
kpc of the disk and then decreases with height (Figure~\ref{figure.fuvnuv_height}).
This suggests that these filamentary
structures exist alongside an overall change in the dust with height.
In contrast, most of the normal galaxies do not have significant
FUV$-$NUV structure in Figure~\ref{figure.fuvnuv_normal} and the
composite FUV$-$NUV color does not appear to change with increasing
height. This suggests that the dust has similar properties between 
2-12\,kpc. 
However, NGC~891 and NGC~5907 have high $S/N$ data and show a modest
decline with height above 5\,kpc. Deeper data for more galaxies are necessary
to determine whether the color is constant with height.
Finally, in Figure~\ref{figure.fuvnuv_height} the stripped
galaxies appear bluer than starburst galaxies at low heights and
redder at larger heights. Since some of the FUV emission comes from
extraplanar star formation in these few cases, we speculate that
the difference is not explained by a change in the dust with height.

If the FUV$-$NUV color change with height around starbursts results
from a change in the dust, two plausible explanations are that
galactic winds eject a different type of dust and carry it to larger
heights than the processes in normal galaxies, or that winds and a
comparatively intense radiation field alters the pre-existing dust at
large heights. In normal galaxies, the
dust may circulate through a galactic fountain \citep{bregman80},
which is consistent with the measured UV scale heights of several
kpc. On larger scales, we expect the dust to change from Milky Way-like dust
in the disk to SMC-like dust by a radius of 20\,kpc \citep{menard10},
which (for a given radiation field) will lead to smaller FUV$-$NUV
color \citep[however, cf.][who find that most ``extragalactic''
  dust can be explained by an extended disk]{smith16}. In the
composite (Figure~\ref{figure.fuvnuv_height}) we do not detect the
transition through FUV$-$NUV color alone, but most of the galaxies do
not have reliable FUV$-$NUV colors above 10\,kpc; the values there for
normal galaxies come from 2-4 galaxies, depending on the height. A
clean test for this transition requires deeper data for more
individual galaxies. If the dust does not change around normal
galaxies up to heights difficult for a galactic fountain to reach, one
possibility is that starburst episodes provide most of the halo dust.

\subsection{M82 and NGC 891}

NGC~891 and M82 both have UV halos that have been previously studied,
and here we compare our maps and results to those in the
literature. NGC~891 is a normal Milky Way analog with an elevated star
formation rate near 4\,$M_{\odot}$\,yr$^{-1}$, \citep{popescu04} and a
bright X-ray halo \citep{bregman94}, whereas M82 is the archetypal
superwind galaxy.

The UV halo around NGC~891 was modeled by \citet{seon14}, who found
that the UV light is consistent with a scale height between 1.2-2\,kpc
and a dust mass above a projected height of 2\,kpc of 3-5\% of the
total. \citet{hughes14} find a dust mass in the disk of
$M_{\text{dust}} = 8.5\times 10^7 M_{\odot}$, so the halo dust mass
from the MCRT models in \citet{seon14} is between $2.5 \times
10^6-4.3\times 10^6 M_{\odot}$. This is consistent with the MCRT scale
height and mass we measured: $h=1.5\pm0.2$\,kpc and 
$M_{\text{dust}} =(6\pm3)\times 10^6 M_{\odot}$ (for Milky Way dust),
whereas \citet{seon14} fit a profile along
the projected minor axis, incorporating the core of the PSF but not
the highly extended wings. 

In \citetalias{hk14}, where we did not account for the extended PSF
contamination of the halo, the UV intensity falls off more slowly with
height on the side with the large \ion{H}{1} filament, which also
occurs in the X-rays and the 160\,$\mu$m image. The combined,
PSF-corrected UV map shows that the part of the UV halo clearly
associated with NGC~891 is brighter on that side
(Figure~\ref{figure.gallery_normal}), but the large-scale very diffuse
emission is too extensive to be scattered light from dust around
NGC~891 and may be Galactic cirrus. The increased UV intensity near
the galaxy may arise from dust within the filament material adding to
the total dust column.

\citet{howk97} examined extraplanar dust around NGC~891 in extinction
at a lower height than the UV halo, finding filamentary structures
that form a thick disk and have a mass of at least a few$\times 10^7
M_{\odot}$, or about 10-50\% of the total \citep{hughes14}. The
resolution of their images is much higher than the UV images, so we
cannot rule out the possibility that the eRN is also highly
structured. However, the mass in the scattering component (estimated
through MCRT or 160\,$\mu$m flux) is less than half than that inferred
by \citet{howk97}. This is also true if we measure the diffuse dust
mass between 0.5-2\,kpc in our MCRT models. While they argue that it
would be challenging for a galactic fountain \citep{bregman80} or
radiation pressure \citep{ferrara91} to lift enough dust mass to
account for the extinction features they see, either mechanism could
produce the smaller amount of diffuse dust inferred at large
heights. This suggests that much of the dust in the chimneys falls
back to the disk, and that the diffuse medium above 2\,kpc is a
separate component.

M82 has one of the first UV halos reported \citep{hoopes05}, since its
extraplanar diffuse UV light is obvious from the image even without
special processing. It is also the best-studied UV halo
\citep{coker13,hutton14}, and the dust content of the wind has also
been investigated by several authors
\citep[e.g.,][]{engelbracht06,roussel10,yoshida11}. The consensus that
the UV halo in M82 is an eRN supports the identification of other UV halos as
eRN.  The FIR dust emission is seen to about the same or larger
distances than in the UV, implying a mass of about $10^6 M_{\odot}$
\citep{roussel10}. It is more difficult to construct an accurate MCRT
model for M82 than other disk galaxies because of its irregular
\ion{H}{1} morphology, but 
using the same method as described above we obtain a value of
$M_{\text{dust}} = 2^{+3}_{-1}\times 10^6 M_{\odot}$, which is consistent
with the FIR measurement. M82 also allows us to test our de-reddening
procedure: our reported FUV $L_{\text{gal}} = 3\times
10^{42}$\,erg\,s$^{-1}$, whereas \citet{coker13} estimate that the UV
luminosity seen by the halo is $L_{\text{gal}} = 1-6\times
10^{42}$\,erg\,s$^{-1}$. 

We suggest that the best simple explanation for the the FUV$-$NUV
color change with height (in M82 and starburst halos generally) is a
change in the dust. We note that \citet{roussel10} show that there is
little change in the 250\,$\mu$m/350\,$\mu$m ratio in the region with
the UV halo, but if the change is primarily one of composition rather
than size the FIR bands may not be sensitive to it. For example,
\citet{nozawa13} successfully fit a two-component (graphite and
silicate) dust model to extinction along different Milky Way and SMC
sight lines by changing the mixture but fixing the size distribution
using the same law found by \citet{hutton14}. Likewise,
\citet{hutton14} found, using shallow \Swift{} data, that the dust is
consistent with a single size distribution. They focused on
color-color plots where a change in composition is degenerate with a
change in the incident spectrum, so neither of these studies presents
strong evidence against a change in the dust type with
height. However, we emphasize that we have not conclusively shown that
the dust composition must change.

\section{Summary and Conclusions}
\label{section.summary}

We have described the morphology of UV halos around nearby, edge-on
galaxies with and without superwinds, and here summarize our findings:
\begin{enumerate}
\item UV halos are astrophysical and can be separated from galactic light
  scattered into the PSF wings (Airy patterns). They are broadband
  phenomena and truly diffuse, with a flux of 1-20\% of the apparent
  galaxy luminosity 
  (a few percent or less of the de-reddened values). 
  They are visible to beyond 10\,kpc above the midplane.
\item Around normal galaxies, UV halos tend to have a thick-disk
  morphology, but they differ in their radial concentration and
  vertical prominence. In superwind galaxies, the UV halos are visible
  around the whole galaxy and contain filamentary structures seen at
  other wavelengths and associated with the winds. Galaxies that are
  being stripped of their ISM by ram pressure have asymmetric UV halos
  and some extraplanar star formation.
\item Among a variety of galactic parameters, we found that the UV
  halo luminosity is only correlated with the galaxy luminosity and
  the SFR.
\item The structure, broadband visibility, and strong correlation of
  $L_{\text{halo}}$ with $L_{\text{gal}}$ for UV halos lead us to
  conclude that they are eRN, which supports prior arguments made in
  the literature. 
\item The frequency of UV halos (100\% in our UV-selected sample of
  highly inclined galaxies within 25\,Mpc) indicates that eRN are
  ubiquitous, but not as extensive as reported in \citetalias{hk14}. 
\item The dust mass of the diffuse component above 2\,kpc, inferred
  from MCRT models that are constrained by the measured
  $L_{\text{halo}}/L_{\text{gal}}$ and 21-cm maps, is a few percent of
  the dust mass in the disk, and perhaps 10\% of the dust mass seen in filamentary
  extraplanar structures at lower heights. Deeper FIR observations are
  needed to constrain $\beta$ in the halo dust for a direct measurement.
\item There is tentative evidence for a change in the dust properties with
  height in starburst galaxies, but normal galaxies are consistent
  with constant dust properties between 2-12\,kpc. 
\end{enumerate}
We expect virtually every star-forming galaxy to form an eRN, but
their detectability in a given sample is a strong function of the UV
luminosity of the galaxy and the inclination. Beyond about 25\,Mpc it
becomes difficult to isolate filamentary structure with \Galex{} or
\Swift{}, but within this limit the UV halos around normal galaxies
appear smooth, and there are no clear features in the FUV$-$NUV color
maps. One consequence is that the height profile used to constrain
MCRT models \citep{seon14,shinn15} provides a reasonable estimate of
the dust mass, assuming a thick disk morphology. UV observations of
more distant starburst or wind galaxies may 
not discern the wind structure, but an unusually extensive or blue UV
halo may indicate the presence of a wind. 

Our results confirm the presence of diffuse dust around galaxies as a
general phenomenon. The dust is a reliable tracer of material that, at
one point, originated in a galaxy disk, so the amount and physical
properties (grain size distribution, chemical composition,
temperature, etc.) tell us about the role of stellar (or possibly AGN)
feedback in the history of the galaxy and the pollution of the
circumgalactic medium with metals. However, much work remains: a
larger, less biased sample of UV halo properties is needed, at the
cost of poorer resolution. This will be addressed in Paper~II. In
Paper~III, we will constrain dust properties in a simple model using
the UV SED. Beyond this, we need to understand the dust outflow
mechanism (the roles of radiation pressure and hydrodynamic
entrainment), connect the UV SED to the FIR SED, and determine if (or
how) the dust within 10\,kpc of the galaxy connects to a larger
circumgalactic component. 

\acknowledgments

We thank the anonymous referee for a helpful report that improved the
quality of this paper.
This research has made use of the NASA/IPAC Extragalactic Database
(NED) which is operated by the Jet Propulsion Laboratory, California
Institute of Technology, under contract with the National Aeronautics
and Space Administration. We acknowledge the usage of the HyperLeda
database (http://leda.univ-lyon1.fr). E.~H.-K. and J.C. gratefully acknowledge
support from NASA grant NNH13ZDA001N-SWIFT. 

{\it Facilities:} \facility{GALEX}, \facility{Swift}


\appendix

Tables~\ref{table.obsIDsMrk501} and \ref{table.obsIDsGalaxies} contain the
\Swift{} and \Galex{} observation IDs for the Mrk~501 PSF measurements and 
the galaxies used in this work, respectively. The \Galex{} PSF models were
based on data sets for 3C~273 (obsID GI4\_012003\_3C273) and PKS~$2155-304$
(obsID PKS2155m304).

\begin{deluxetable*}{ccc}
\tablenum{9}
\tabletypesize{\scriptsize}
\tablecaption{\Swift{} Mrk 501 Observations}
\tablewidth{0pt}
\tablehead{
\multicolumn{3}{c}{ObsIDs} \\
\colhead{\uvmtwo{}} & \colhead{\uvwone{}} & \colhead{\uvwtwo{}}
}
\startdata
30793006 &	30793001 &	30793006 \\
30793007 &	30793002 &	30793007 \\
30793008 &	30793003 &	30793008 \\
30793009 &	30793004 &	30793009 \\
30793010 &	30793006 & 	30793010 \\
30793011 &	30793007 &	30793011 \\
30793012 &	30793008 & 	30793012 \\
30793013 &	30793009 & 	30793013 \\
30793014 &	30793010 & 	30793014 \\
30793015 &	30793011 & 	30793015 \\
30793016 &	30793012 & 	30793016 \\
...      &  ...      &  ...       
\enddata
\tablecomments{\label{table.obsIDsMrk501} \Swift{} observations of Mrk~501
used to construct the UVOT PSF models used in this paper. The full table is
available online.}
\end{deluxetable*}

\begin{deluxetable*}{lccccc}
\tablenum{10}
\tabletypesize{\scriptsize}
\tablecaption{\Swift{} Mrk 501 Observations}
\tablewidth{0pt}
\tablehead{
\colhead{Galaxy} & \multicolumn{5}{c}{ObsIDs} \\
                 & \colhead{FUV} & \colhead{NUV} & \colhead{\uvwtwo{}} & \colhead{\uvmtwo{}} & \colhead{\uvwone{}} 
}
\startdata
M82		& GI1\_071001\_M81 &	GI1\_071001\_M81 & 31201001 &	31201001 & 31201001 \\
		&				 & 				   & 32503004 &	31201002 & 32503002 \\
		&				 & 				   & 32503016 &	32503003 & 32503015 \\
		&				 & 				   & 32503020 &	32503007 & 32503018 \\
		&				 & 				   & 32503028 &	32503008 & 32503022 \\
		&				 & 				   & 32503032 &	32503011 & 32503025 \\
		&				 & 				   & 32503036 &	32503017 & 32503026 \\
		&				 & 				   & 32503040 &	32503021 & 32503034 \\
		&				 & 				   & 32503046 &	32503024 & 32503043 \\
		&				 & 				   & 32503057 &	32503029 & 32503045 \\
...     & ... 			 & ...			   & ...	  & ...      & ... 
\enddata
\tablecomments{\label{table.obsIDsGalaxies} \Swift{} and \Galex{} observations 
used in this paper. The full table is available online.}
\end{deluxetable*}

\end{document}